\documentclass[onecolumn,usenatbib]{mn2e}
\usepackage{natbib}
\usepackage{amsmath}
\usepackage{graphicx}
\def\beq{\begin{eqnarray}}
\def\eeq{\end{eqnarray}}

\usepackage{amsbsy}
\usepackage{epsf}
\def\X{{\mathrm{x}}}
\def\Y{{\mathrm{y}}}

\def\n{{\rm n}}
\def\p{{\rm p}}

\def\bep{\bar{\varepsilon}}
\def\bB{\bar{\mathcal{B}}}
\def\bBp{\bar{\mathcal{B}}'}

\newcommand{\be}{\begin{equation}}
\newcommand{\ee}{\end{equation}}
\newcommand{\bear}{\begin{eqnarray}}
\newcommand{\eear}{\end{eqnarray}}

\begin{document}

\title{r-modes and mutual friction
in rapidly rotating superfluid neutron stars}

\author[B. Haskell et al]
{B. Haskell, N. Andersson, A. Passamonti \\
School of Mathematics, University of Southampton,
Southampton, SO17 1BJ, United Kingdom\\}

\maketitle

\begin{abstract}
We develop a new perturbative framework for studying the r-modes of rotating superfluid
neutron stars. Our analysis accounts for the centrifugal deformation of the star, and considers
the two-fluid dynamics at linear order in the perturbed velocities. Our main focus is on a simple model
system where the total density profile is that of an $n=1$ polytrope. We derive a partially analytic solution
for the superfluid analogue of the classical r-mode. This solution is used to analyse the relevance of the
vortex mediated mutual friction damping, confirming that this dissipation mechanism is unlikely
to suppress the gravitational-wave driven instability in rapidly spinning superfluid neutron stars.
Our calculation of the superfluid r-modes is significantly simpler than previous approaches,
because it decouples the r-mode from all other inertial modes of the system. This leads to the
results being clearer, but it also means that we cannot comment on the relevance of
potential avoided crossings (and associated ``resonances") that may occur for particular parameter values.
Our analysis of the mutual friction damping differs from previous studies in two important ways.
Firstly, we incorporate realistic pairing gaps which means that the regions of superfluidity in the star's core
vary with temperature.
Secondly, we allow the mutual friction parameters to take the whole range of permissible values rather than
focussing on a particular mechanism. Thus, we consider not only the weak drag regime, but also the strong
drag regime where the fluid dynamics is significantly different.
\end{abstract}

\maketitle

\section{Introduction}

Oscillations of rapidly rotating neutron stars have attracted interest for a considerable time.
During the last decade significant effort was aimed at understanding whether gravitational-wave emission
sets the upper speed limit for pulsars, e.g. via the r-mode instability \citep{Nils1998}.
This possibility is of particular interest since such unstable systems may radiate detectable gravitational
waves. In particular, it was suggested that r-modes in rapidly rotating neutron stars in low-mass X-ray binaries (LMXBs)
may lead to a persistent source of gravitational radiation \citep{Bildsten1998, Nils1999} that may be detected by advanced LIGO [see \citet{Anna} for the most recent analysis of this problem].
It is of great importance to understand whether internal fluid dissipation  allows the instability to develop in such systems or whether it suppresses the r-modes completely.

There are, in fact, many mechanisms at work in a real neutron star which compete with the gravitational-wave driving of the r-mode. One can express the
relative strength of these mechanisms in terms of timescales which will, in general, depend on a large number of parameters. For a basic instability analysis, the most important parameters are
 the rotation rate and the temperature of the star. The instability can only develop when the gravitational radiation growth time-scale is shorter than the damping timescales due to the various viscosity mechanisms. This defines a region in the spin-temperature parameter space where the r-mode instability is active.
Several studies have been devoted to calculating the shear- and bulk viscosity timescales for more or less realistic neutron star models. These studies tend to agree that the notion of persistent gravitational-wave emission accreting neutron stars in LMXBs is quite robust [see \citet{Owen2006} and references therein].

The neutron stars in LMXBs are believed to be old cold recycled pulsars that have been spun up by accretion. These stars are expected to have superfluid cores and are thus expected to have rather different fluid dynamics and dissipation mechanisms. The simplest description of these systems is provided by a two-fluid model, where the superfluid neutrons are distinguished from the proton-electron plasma (usually treated as a charge neutral fluid).
The oscillations of neutron stars with such superfluid cores have been studied by \citet{Lind1} and \citet{NilsGreg2001}. The latter showed that there are no g-modes in such stars, but on the other hand one has a new
class of ``superfluid'' modes. In particular, there are two families of r-modes \citep{prix}. One of these resembles the ordinary r-modes of a barotropic star in the sense that the neutrons and protons ``move together". For the other set of modes the neutrons and protons are ``counter-moving". The relative motion of neutrons and protons in a superfluid is, however, rapidly damped by mutual friction, a mechanism that is directly associated with the rotational neutron vortices.
The most commonly considered mutual friction mechanism is
the scattering of electrons off the magnetic fields associated with the superfluid neutron vortices \citep{als,trev}. Mutual friction is known to have a decisive impact on neutron star dynamics. In fact,
\citet{Lind2} have argued that mutual friction completely suppresses the gravitational radiation driven f-mode instability [see \citet{fmode} for a recent discussion]. It is thus important to understand if the r-mode instability suffers a similar fate. Previous work on the subject by \cite{Lind3} and \cite{yl2} presents a somewhat mixed picture. While the ordinary r-mode is very weakly damped for most values of the entrainment parameter, the damping becomes very strong for a small range of parameter values. Hence, there may be situations where the mutual friction is strong enough to stop the instability from growing.

The purpose of this paper is to clarify the role of the mutual friction damping and understand whether it can stabilize the r-modes against the gravitational-wave instability. In order to do this we extend the formulation of \citet{fmode} to second order in rotation to calculate the mutual damping timescale for r-modes. Although we do not expect our analysis to lead to results that change  previous conclusions,  there are good reasons to return to this problem. First of all, previous studies have focussed entirely on the weak drag regime for the mutual friction. Meanwhile, recent studies of superfluid turbulence \citep{turb,peralta05,peralta06}, neutron star free precession \citep{letter,prec2} and pulsar glitches \citep{glitch} demonstrate that systems in the strong drag regime may have very dynamics.
Since one can make convincing arguments for the strong drag regime being relevant \citep{strong1,strong2,strong3,strong4}, we clearly need
to understand the r-mode problem for this parameter range as well. Secondly, we want to improve the treatment of
the critical temperature/density at which superfluidity comes into play. The density dependence of the various superfluid pairing gaps translates into distinct regions of normal- and super-fluids that vary with the
core temperature. So far, the best analysis in this respect is that of \citet{Lind3} who consider a two-fluid core
surrounded by a single fluid envelope. We will build on this by considering superfluid layers, the size of which are
determined by a (qualitatively) realistic pairing gap. The size of the superfluid region affects not only the
mutual friction but also the shear- and bulk viscosities, and it is relevant to demonstrate how this
alters the r-mode instability window. Finally, we want to lay the foundation for more detailed work on exotic neutron star cores.
It is well established that exotic cores, dominated by either hyperons or deconfined quarks, may be associated with a very strong bulk viscosity [see, for example, \citet{Owen2006}].
Although it has been suggested that this would lead to the suppression of the r-modes, it is clear that such a conclusion is premature. In reality one would expect
superfluidity to play a role. If the hyperons are superfluid the reactions that
produce the bulk viscosity are suppressed and the effect on the r-mode instability may not be that severe \citep{Owen2006}.
A similar argument applies to colour superconducting quarks \citep{Alford1999}. However, it may be important to keep in mind that a superfluid system has extra
dynamical degrees of freedom. In order to truly understand the dynamics of these exotic phases of matter we need to explore the multi-fluid nature of these systems. The present study paves the way for
future work in this direction.

\section{The two-fluid equations}

\subsection{The unperturbed problem}

Our discussion is based on the usual two-fluid model for neutron star cores \citep{variation,monster}. That is,
we consider two dynamical degrees of freedom, loosely speaking representing the superfluid neutrons
(labeled $\n$) and a charge-neutral conglomerate of the protons and electrons (labeled $\p$).
Assuming that the individual species are conserved, we have the
standard conservation laws
\be
\partial_t \rho_\X + \nabla_i (\rho_\X v_\X^i) = 0\label{continuo1}
\ee
where the constituent index x may be either p or n. The equations of momentum balance can be written
\be
(\partial_t + v_\X^j \nabla_j ) (v^\X_i+\varepsilon_\X w^{\Y\X}_i) +\nabla_i (\tilde{\mu}_\X+\Phi)
+ \varepsilon_\X w^j_{\Y\X} \nabla_i v^\X_j= f^\X_i/\rho_\X
\label{Eulers}\ee
where $w_i^{\Y\X}=v_i^\Y-v_i^\X$ ($\Y\neq\X$), and $\tilde{\mu}_\X=\mu_\X/m_\X$ represents the
chemical potential (in the following we assume that $m_\p=m_\n$). Moreover, $\Phi$ represents the gravitational potential, and the parameter
$\varepsilon_\X$ encodes the  entrainment effect.
The force on the right-hand side
of (\ref{Eulers}) can be used to represent  other interactions, including dissipative terms \citep{monster}. We will focus on the 
vortex-mediated mutual friction force for a system that,  in equilibrium, rotates uniformly. This means that we
consider a force of form \citep{trev}
\be
f^\X_i = 2  \rho_\n \mathcal{B}' \epsilon_{ijk}\Omega^j w_{\X\Y}^k
+ 2  \rho_\n  \mathcal{B} \epsilon_{ijk}\hat{\Omega}^j \epsilon^{klm} \Omega_l w_m^{\X\Y}
\label{mf}\ee
Here, $\Omega^j$ is the angular frequency of the neutron fluid (a hat represents a unit vector).

One can express the mutual friction force in terms of a dimensionless "drag" parameter $\mathcal R$
such that \citep{trev}
\be
\mathcal{B} = { \mathcal{R} \over 1 + \mathcal{R}^2} \ , \qquad \mbox{ and } \qquad
\mathcal{B}' = { \mathcal{R}^2 \over 1 + \mathcal{R}^2}
\ee
In the standard picture, the mutual friction is due to the scattering of electrons off of an array of neutron vortices \citep{als} which leads to $\mathcal{R}\ll 1$, thus placing the problem in the weak drag regime. This means that $\mathcal{B}\ll 1$ and $\mathcal{B}' \ll \mathcal{B}$.
Hence, all $\mathcal{B}'$ terms can be neglected. However, it is commonly thought \citep{strong1,strong2,strong3} that
the strong drag limit, $\mathcal{R} \gg 1$, may apply. Intermediate values for the drag, $\mathcal{R}\approx 1$ are also interesting. In particular since  one would then have both $\mathcal{B}\approx 1$ and $\mathcal{B}' \approx 1$ (note that in the case $\mathcal{R}\gg 1$ one still has $\mathcal{B}' \approx 1$ but $\mathcal{B}\ll 1$) . This leads to the presence of terms that may have significant effect on the dynamics of the system
\citep{letter,prec2,glitch}. At this point, our understanding of neutron star core physics is sufficiently rudimentary
that we should avoid ruling out the various possibilities. Hence, we will consider  the entire permissible range of values for the drag
parameter.

Let us consider a frame rotating with the star at fixed angular velocity $\Omega^i$. The equations of motion then take the form:
\be
(\partial_t + v_\X^j \nabla_j ) (v^\X_i+\varepsilon_\X w^{\Y\X}_i) +2 \epsilon_{ijk}\Omega^j  v_\X^k +\nabla_i (\tilde{\mu}_\X+\Phi_R)
+ \varepsilon_\X w^j_{\Y\X} \nabla_i v^\X_j= f_i^\X / \rho_\X
\label{Eulers2}\ee
where we have included the centrifugal term in the potential
\be
\Phi_R=\Phi-\frac{1}{2}\Omega^2r^2\sin^2\theta\label{potrot}
\ee
The continuity equations maintain the  form (\ref{continuo1}) and the Poisson equation is
\be
\nabla^2 \Phi = 4\pi G \sum_\X \rho_\X
\ee

\subsection{Perturbations}

To keep the problem tractable we will
assume that the background configuration is such that the two fluids rotate together [see \citet{prix,glitch} for discussions
of oscillations in systems that are not in co-rotation]. Perturbing the equations of motion and working in a frame rotating with $\Omega^j$ we then have
\be
\partial_t (\delta v^\X_i + \varepsilon_\X\delta w_i^{\Y\X}) + \nabla_i (\delta \tilde{\mu}_\X + \delta \Phi)
+ 2 \epsilon_{ijk}\Omega^j \delta v_\X^k =  \delta ( f^\X_i/\rho_\X)
\ee
and
\be
\partial_t \delta \rho_\X + \nabla_j (\rho_\X \delta v_\X^j) = 0
\ee

To completely specify the perturbation problem, we need boundary conditions. At the centre of the star
we simply require that all variables are regular. The surface of the star is somewhat more complex. In reality one
does not expect the superfluid region to extend all the way to the surface.  We shall thus assume that the superfluid is only present
in a distinct region determined by the core temperature and the superfluid pairing gap. We will discuss the  implementation of this idea
 once we have set up the relevant system of equations.

From previous work on superfluid neutron star oscillations, e.g. \citet{Lind1,NilsGreg2001,rieut,fmode},
we know that the problem has two "natural" degrees of freedom. One of them represents the total mass flux.
Introducing
\be
\rho \delta v^j = \rho_\n \delta v_\n^j + \rho_\p \delta v_\p^j
\ee
and combining the two Euler equations accordingly, we get
\be
\partial_t \delta v^j +  \nabla_i \delta \Phi + { 1 \over \rho} \nabla_i \delta p- \frac{\delta \rho}{\rho^2}\nabla_i p
+ 2 \epsilon_{ijk}\Omega^j \delta v^k = 0
\label{Eulav}\ee
where $\rho = \rho_\n+\rho_\p$ and the pressure is obtained from
\be
\nabla_i p = \rho_\n \nabla_i \tilde{\mu}_\n + \rho_\p \nabla_i \tilde{\mu}_\p
\ee
We also have
\be
\partial_t \delta \rho + \nabla_j (\rho \delta v^j) = 0
\label {conav}\ee

At this point we have two equations which are identical to the perturbation equations for a single fluid
system. Of course, we are considering a two-fluid problem and there is a second dynamical
degree of freedom. To describe this, it is natural to consider the difference in velocity. Thus, we introduce
\be
\delta w^j = \delta v_\p^j - \delta v_\n^j
\ee
Combining the two Euler equations in the relevant way we have
\be
(1-\bep) \partial_t \delta w_i + \nabla_i \delta \beta + 2 \bBp \epsilon_{ijk}\Omega^j \delta w^k - 2\bB \epsilon_{ijk}\hat{\Omega}^j \epsilon^{klm} \Omega_l \delta w_m = 0
\label{Euldi}\ee
We have defined
\be
\delta \beta = \delta \tilde{\mu}_\p - \delta \tilde{\mu}_\n
\ee
which represents the (local) deviation from chemical equilibrium induced by the perturbations. We have also introduced
\be
\bep = \varepsilon_\n/x_\p \ , \qquad \bBp = 1-\mathcal{B}'/x_\p \ , \qquad
\bB = \mathcal{B}/x_\p
\ee
and we remind the reader that
\be
\rho_\p\varepsilon_\p=\rho_\n\varepsilon_\n
\ee
Finally, we need a second "continuity" equation. To close the system, it  seems natural to consider an
equation for the proton fraction $x_\p$. We then find that
\be
\partial_t \delta x_\p + { 1 \over \rho} \nabla_j \left[ x_\p (1-x_\p) \rho \delta w^j \right] +  \delta v^j \nabla_j x_\p = 0
\label{condi}\ee

\subsection{Model equation of state}

Let us now consider the set of equations that we have written down. We see that the two
degrees of freedom $[\delta v_i, \delta p]$ and $[\delta w_i, \delta \beta]$ only couple ``directly" through
(\ref{condi}).
For compressible models, the two degrees of freedom also couple ``indirectly", since we can use the equation
of state to relate $[\delta p,\delta \beta]$  to $[\delta \rho, \delta x_\p]$.

Deciding to work with $\delta p$ and $\delta \beta$ [cf. \citet{fmode}], we use
\be
\delta\rho=\left(\frac{\partial\rho}{\partial p}\right)_\beta\delta p + \left(\frac{\partial\rho}{\partial\beta}\right)_p\delta\beta
\ee
and
\be
\delta x_\p= \left(\frac{\partial x_\p}{\partial p}\right)_\beta \delta p + \left(\frac{\partial x_\p}{\partial\beta}\right)_p \delta\beta
\ee
As a simple model,
we shall consider an equation of state such that
\be
\left(\frac{\partial\rho}{\partial p}\right)=\frac{1}{c_s^2}
\ee
where $c_s$ is the background sound speed. We combine this with a simple linear expression for the proton fraction;
\be
x_\p=\alpha\rho
\ee
This leads to
\be
\left(\frac{\partial x_\p}{\partial p}\right)=\frac{\alpha}{c_s^2}\;\;\;\mbox{ and }\;\;\;\left(\frac{\partial x_\p}{\partial\beta}\right)=\frac{\alpha^2\rho^2}{c_s^2}
\ee
where we have used the relation  \citep{NilsGreg2001}
\be
\left(\frac{\partial\rho}{\partial\beta}\right)=\rho^2\left(\frac{\partial x_\p}{\partial p}\right)
\ee
 to get
\be
\left(\frac{\partial\rho}{\partial\beta}\right)=\frac{\alpha\rho^2}{c_s^2}
 \ee
In our model calculations we shall take $\alpha=6\times 10^{-3} / \rho_\mathrm{nuc}$ where $\rho_\mathrm{nuc}\approx 2.8\times 10^{14}$ g/cm$^3$ is the nuclear saturation density.

In a superfluid system the momentum of each component may not be parallel to it's velocity. Rather, it  acquires a component along the relative velocity. This is evident from equation (\ref{Eulers}).
This non-dissipative coupling is usually parametrised in terms of the entrainment parameter $\varepsilon_\mathrm{x}$.
One can show \citep{prix} that this parameter is linked to the effective proton mass, $m_\p^*$, according to
\be
\varepsilon_\p=1-\frac{m_\p^*}{m_\p}
\ee
For simplicity, we shall take $\varepsilon_\p$ to be constant throughout the superfluid region of the
star. Recent work suggests that, while this may not be a good approximation for an entire neutron star, it is approximately true if we consider a shell.
From \citet{chamel} we see that a reasonable range for the entrainment parameter is $\varepsilon_\p\approx 0.2-0.8$ 

In order to compare our results to previous work, it is worth recalling that the entrainment
parameter $\epsilon$ used by \cite{Lind3} is related to $\varepsilon_\p$ by
 \be
 \epsilon=\frac{\varepsilon_\p x_\p}{1-x_\p-\varepsilon_\p}
 \ee
Hence,  their range of $\epsilon\approx 0.02-0.06$ corresponds, if we take a volume average of $x_\p$, to $\varepsilon_\p\approx 0.45-0.85$.

To summarise; When we solve the r-mode problem, we will consider an equation of state represented by i) the overall
density profile, represented by the
 sound speed  $c_s^2$, ii) the proton fraction $x_\p$ and iii) the entrainment parameter $\varepsilon_\p$.
These quantities allow us to specify the background needed for the perturbation problem.
This model is quite simplistic, and it would not be difficult to make it
more realistic. However, our main interest is to explore how the r-mode results depend on
the different parameters. This question is easier to address with a simple parametrised model.

\section{Slow rotation analysis}
\label{slowrot}

In order to determine the rotational corrections to the superfluid r-modes, we will apply
the formalism of \cite{Saio1982} to the problem. Thus, we consider corrections up to order $O(\Omega^2)$ in the slow rotation approximation. We will  also make the Cowling approximation, i.e. neglect perturbations of the gravitational potential, $\delta\Phi$.

We start by considering a slowly, and uniformly, rotating star. It is well known that such a star will not be spherical and that a distorted potential surface can be written in the form
\be
r=a[1+\epsilon(a,\theta)]
\label{variabile}\ee
Here $\epsilon$ is a function of $a$ and $\theta$ which represents the deformation of the equilibrium structure from the background spherical state. The equilibrium physical quantities are functions only of $a$. Following \citet{Saio1982,Smeyers}, we  write the equations of motion in a frame corotating with the star, denoted by $\{q^i\}$, starting from a static Cartesian frame, denoted by $\{x^i\}$.
Our coordinates in the rotating frame will be a spherical polar system, explicitly:
\bear
x^1&=&a(1+\epsilon)\sin\theta\cos(\phi+\Omega t)\\
x^2&=&a(1+\epsilon)\sin\theta\sin(\phi+\Omega t)\\
x^3&=&a(1+\epsilon)\cos\theta
\eear
The metric in the new coordinates is then given by
\be
g_{ab}=\delta_{ij}\frac{\partial x^i}{\partial q^a}\frac{\partial x^j}{\partial q^b}
\ee
which, after linearizing with respect to $\epsilon$, leads to
\be
g_{ab}=\left\{\begin{array}{c c c}
1+2\epsilon+2a\frac{\partial\epsilon}{\partial a} & a\frac{\partial\epsilon}{\partial \theta} & 0\\
a\frac{\partial\epsilon}{\partial \theta} & a^2(1+2\epsilon) & 0\\
0 & 0 & a^2(1+2\epsilon)\sin^2\theta \end{array}\right\}+O(\epsilon^2)
\ee
The connection coefficients, in the coordinate basis
\be
 \left\{e_a=\frac{\partial}{\partial a}, e_\theta=\frac{\partial}{\partial\theta}, e_\phi=\frac{\partial}{\partial\phi}\right\}
\ee
can be obtained from
\be
\Gamma^a_{bc}=\frac{1}{2}g^{ad}\left(g_{bd,c}+g_{cd,b}-g_{bc,d}\right)
\ee
We shall, however, use the vector basis
\be
\left\{\hat{e}_i\right\} =\left\{\hat{e}_a=\frac{\partial}{\partial a}, \hat{e}_\theta=\frac{1}{a}\frac{\partial}{\partial\theta}, \hat{e}_\phi=\frac{1}{a\sin\theta}\frac{\partial}{\partial\phi}\right\}
\ee
for which the covector basis is
\be
\left\{\hat{\omega}^i\right\}=\left\{da,ad\theta,a\sin\theta d\phi\right\}
\label{base}\ee
Note that the basis vectors are not orthogonal.

Before we proceed, it is worth recalling that  \cite{Smeyers} have argued that the derivation of Saio's results is flawed, even though the final results for the
rotational frequency correction
are correct.
Since this is an important point, we will outline how the second order slow-rotation perturbation equations
should be derived following Saio's strategy.

As we are considering linear perturbations on a rotationally distorted background, the first step is to calculate such a background, using the coordinates $(a,\theta,\phi)$ defined above.
The equations that govern the background equilibrium are particularly simple (if we assume that neutrons and protons do not move relative to each other)
\be
\frac{1}{\rho(a)}\frac{dP(a)}{da}=-\frac{d\Phi_R(a)}{da}
\ee
where $\Phi_R$,  defined in equation (\ref{potrot}), includes the centrifugal terms.
This is due to the fact that the coordinate $a$ has been defined in such a way as to label the equipotential surfaces of $\Phi_R$, which are also isobaric (and isopycnic as the background
equation of state is barotropic) surfaces of the star, thus leading to all the background quantities being a function of $a$ only [for a detailed description see \cite{rotating}].
The Euler equations are thus identical to those one would obtain for a spherical background and the solution can be formally obtained by replacing the radial coordinate $r$ with the variable $a$ in the spherically symmetric solution. It is important to remember, however, that we are calculating physical quantities at a point $P$ in the deformed star labeled by the coordinates $(a,\theta,\phi)$ and not at a point $P_0$ in the spherical star labeled by coordinates $(r=a,\theta,\phi)$.
Note, in fact, that as the geometry is not spherical the measures of distance and volume change. For example, the mass of the star is now obtained from
\be
M=\int_{a=0}^R\int_{\theta=0}^\pi\int_{\phi=0}^{2\pi} \rho(a)a^2\sin \theta \left[1+3\epsilon(a,\theta)+a\frac{d\epsilon(a,\theta)}{da}\right] da\;d\theta\; d\phi=M_0+\delta M_{\Omega}
\label{masse}\ee
where $M_0$ is the mass of the spherically symmetric star and $\delta M_{\Omega}$ is the difference in mass due to rotation.

\subsection{Perturbations}
\label{slowpert}

Let us now consider linear perturbations on our rotationally deformed background.
We shall express the perturbations in terms of a total displacement vector $\xi^{+}_i$, such that $\delta v_i=i\sigma\xi^{+}_i$, and a difference displacement vector $\xi^{-}_i$, such that $\delta w_i=i\sigma\xi_i^{-}$.
We can then write equation (\ref{Eulav}) as:

\be
\sigma^2\xi^{+}_i-\frac{1}{\rho}\nabla^0_i \delta P+\frac{\delta\rho}{\rho^2}\delta_{ia}\frac{dP}{da}+i\sigma C^{+}_i=0
\label{eulerB}\ee
where the $\delta$ that denotes the Eulerian perturbations should not be confused with the Kronecker delta $\delta_{ia}$.
We are using $i\sigma C^{+}_i$ to represent the Coriolis force, and the vector $C^{\pm}_i$ has the following components:
\bear
C^{\pm}_a\!&=&\!\!2\Omega\left(1+2\epsilon+a\frac{\partial\epsilon}{\partial a}\right)\sin\theta\xi_{\pm}^{\phi}\\
C^{\pm}_\theta\!&=&\!\!2\Omega\left(1+2\epsilon+\tan\theta\frac{\partial\epsilon}{\partial a}\right)\cos\theta\xi_{\pm}^{\phi}\\
C^{\pm}_{\phi}\!&=&\!\!2\Omega\!\left[\xi_\pm^a\sin\theta\left(1+2\epsilon+a\frac{\partial\epsilon}{\partial a}\right)\!+\!\xi_{\pm}^\theta\left(1+2\epsilon+\tan\theta\frac{\partial\epsilon}{\partial\theta}\right)\right]
\eear
The differential operator $\nabla^0_i$ is the same as for spherical polar coordinates, i.e.
\be
\nabla^0_i\xi^i=\frac{1}{a^2}\frac{\partial(a^2\xi^a)}{\partial a}+\frac{1}{a\sin\theta}\frac{\partial(\sin\theta \xi^{\theta})}{\partial\theta}+\frac{1}{a\sin\theta}\frac{\partial\xi^{\phi}}{\partial\phi}
\ee

The continuity equation (\ref{conav}) becomes
\be
\delta\rho+\nabla_0^i\left(\rho\xi^+_i\right)+\rho\xi^+_i\nabla_0^i\left(3\epsilon+a\frac{\partial\epsilon}{\partial a}\right)=0
\label{continuityB}\ee

We shall also need the linearized equation of state. In our model case, we have
\be
\Gamma = { c_s^2 p \over \rho}
\ee
That is, in comparing our equations to Saio's results one should only consider the barotropic limit.

Let us now consider the equations that govern  the second degree of freedom, the difference in velocity.
Equation (\ref{Euldi}) takes the form
\be
\sigma^2(1-\bar{\varepsilon})\xi^-_i-\nabla^0_{i}\delta\beta+i\sigma \bar{\mathcal{B}}^{'} C^{-}_i+i\sigma \bar{\mathcal{B}} G_i=0
\ee
where $G_i$ has the following components:
\bear
G_a\!&=&\!\!\Omega\left[\left(1+2\epsilon+2a\frac{\partial\epsilon}{\partial a}\right)\sin\theta^2\xi_{-}^{a}\right.\nonumber\\
&&\left.+\left(1+2\epsilon+a\frac{\partial\epsilon}{\partial a}+\tan\theta\frac{\partial\epsilon}{\partial\theta}\right)\cos\theta\sin\theta\xi_{-}^{\theta}\right]\\
G_\theta\!&=&\!\!\Omega\left[\left(1+2\epsilon+2\tan\theta\frac{\partial\epsilon}{\partial \theta}\right)\cos\theta^2\xi_{-}^{\theta}\right.\nonumber\\
&&\left.+\left(1+2\epsilon+a\frac{\partial\epsilon}{\partial a}+\frac{\partial\epsilon}{\partial\theta}-\cot{\theta}\frac{\partial\epsilon}{\partial\theta}\right)\cos\theta\sin\theta\xi_{-}^{a}\right]\\
G_{\phi}\!&=&\!\!\Omega\!\left(1+2\epsilon\right)\xi_-^{\phi}
\eear
Finally, we find that
the second continuity equation (\ref{condi}) takes the form
\be
\delta x_\p+\frac{1}{\rho}\nabla_0^i\left[x_\p(1-x_\p)\rho\xi^-_i\right]+{x_\p(1-x_\p)}\xi^-_i\nabla_0^i\left(3\epsilon+a\frac{\partial\epsilon}{\partial a}\right)+\xi^+_i\nabla_0^i x_\p=0
\label{continuityD}\ee

\section{The r-mode problem}

We are mainly interested in understanding the effect that mutual friction has on the r-mode instability. It is
well-known that the r-modes have zero frequency and are purely axial in the non-rotating limit. Rotation breaks this degeneracy and leads to modes that are purely axial to leading order but which have a poloidal component of order $O(\Omega^2)$.
These modes have frequency
\be
\sigma_r=\sigma_0\Omega+\sigma_2\Omega^3
\label{freqord} \ee
The r-mode frequency can be calculated, to second order in rotation, for a single fluid star using a number of different formalisms
\citep{aks,LindOwen,yl}. The superfluid problem has been solved both to linear \citep{NilsGreg2001,prix,fmode} and second \citep{Lind3,yl2} order in rotation.
To leading order one finds that the ordinary r-mode has frequency
\be
\sigma_0=\frac{2m\Omega}{l(l+1)}
\label{rfreq}\ee
where, in fact, only the $l=m$ modes exist.
In addition, a constant density model supports a set of purely axial counter-moving modes with frequency
\citep{fmode}
\be
\sigma_0=\frac{2m\Omega}{l(l+1)(1-\bar{\varepsilon})}\left\{\bar{\mathcal{B}}^{'}+i\bar{\mathcal{B}}\frac{[l(l+1)-m^2]}{m}\right\}\label{s-freq}
\ee
 Since the frequency in (\ref{s-freq}) contains the term $(1-\bar{\varepsilon})$, it is easy to see
 that it can only be the solution for a global mode if $\bar{\varepsilon}=\varepsilon_\p/(1-x_\p)$ is constant. In the present analysis we will assume that
 $\varepsilon_\p$  is constant. Hence, the second class of r-modes can only  exist if we also assume that
the proton fraction $x_\p$ is constant. If $x_\p$ is constant, the co-rotating and counter-rotating degrees of freedom are, in fact, completely decoupled also at second order in rotation. 
The r-mode  is then exactly the same as for a barotropic single fluid
 star. In particular, r-mode solutions exist only for $l=m$. However, we are not generally considering a constant $x_\p$ (in fact we are assuming that the proton fraction scales linearly with the
total mass density, which is not constant). Hence, we see there will be no counter-moving r-modes in our model.
In fact, if $x_\p$ is not a constant, then the coupling between the degrees of freedom leads to the counter-moving
r-mode becoming a general inertial mode, with a mixed toroidal/poloidal velocity field to leading order. These modes
have been determined numerically by \cite{yl2}.
 If we  consider a generic profile for $x_\p$ and work at second order in rotation the leading order
 r-mode will drive the counter-moving degrees of freedom, leading to mutual friction damping. This is the 
effect that we are interested in.

\subsection{Perturbation equations}
\label{rpert}

So far, we have written down the equations for a general perturbation problem on the rotationally deformed background.
We will now focus on modes that are purely toroidal to leading order. That is, we
write the total displacement vector as
\be
\frac{\xi_+^i}{a}=\left(0,\frac{K_{lm}}{\sin\theta}\frac{\partial}{\partial\phi}, -K_{lm}\frac{\partial}{\partial\theta}\right)Y^m_l+\sum_{\nu,\mu}\left(S_{\nu\mu},Z_{\nu\mu}\frac{\partial}{\partial\theta},\frac{Z_{\nu\mu}}{\sin{\theta}}\frac{\partial}{\partial\phi}\right)Y^\mu_\nu
\ee
Analogously, the difference displacement vector can be written
\be
\frac{\xi_-^i}{a}=\left(0,\frac{k_{lm}}{\sin\theta}\frac{\partial}{\partial\phi}, -k_{lm}\frac{\partial}{\partial\theta}\right)Y^m_l+\sum_{\nu,\mu}\left(s_{\nu\mu},z_{\nu\mu}\frac{\partial}{\partial\theta},\frac{z_{\nu\mu}}{\sin{\theta}}\frac{\partial}{\partial\phi}\right)Y^\mu_\nu
\ee
Note that we use uppercase variables for the co-moving degree of freedom and the corresponding lowercase variable for the counter-moving
degree of freedom.

As noted by \cite{Smeyers}, the above form for the displacement vectors does not correspond to the usual decomposition in spheroidal and toroidal components.
At second order in rotation, the above basis differs from the standard basis on a spherical background as the vector $\hat{e}_a$ is no longer orthogonal to $\hat{e}_\theta$.
This is, however, not a problem as we do not need to explicitly identify the poloidal and toroidal parts of the mode at second order in rotation.
Such an identification would not be very useful anyway since the components are coupled at $O(\Omega^2)$.

We want to study the classical r-mode, i.e. a mode that to first order in rotation involves only the co-moving degree of freedom and which is purely toroidal in the slow-rotation limit.
 This identification is unique since our decomposition coincides with the standard one into poloidal and toroidal components  at $O(\Omega)$.
The requirement is  that the leading term $K_{lm}$ is of order unity while the amplitudes of the "spheroidal" components are of order $\Omega^2$  for the total displacement.
Meanwhile, following the first order results of \cite{fmode}, all components of the difference
displacement must be of order $\Omega^2$. As discussed in Appendix~\ref{AppA}, this ordering for the displacement vector, together with the frequency from equation (\ref{freqord}), leads to the following equations
for the $l=m$ r-modes;
\bear
a\frac{dS_{l+1}}{da}&=&\left(\frac{V}{\Gamma}-3\right)S_{l+1}-\frac{V}{\Gamma}\zeta_{l+1}+(l+1)(l+2)Z_{l+1}+3imQ_{l+1}
K_{lm}\left(3D_2+a\frac{dD_2}{da}\right)\label{esse}\\
a\frac{d\zeta_{l+1}}{da}&=&(1-U)\zeta_{l+1}-\frac{V}{\Gamma}x_\p\tau_{l+1}-2ic_1\omega\tilde{\omega} l Q_{l+1}K_{lm\label{dz}}\\
a\frac{ds_{l+1}}{da}&=&\left(X-3\right)s_{l+1}-\zeta_{l+1}\frac{1}{x_\p(1-x_\p)}\frac{V}{\Gamma}+(l+1)(l+2)z_{l+1}-CS_{l+1}+3imQ_{l+1}k_{lm}\left(3D_2+a\frac{dD_2}{da}\right)\\
a\frac{d \tau_{l+1}}{da}&=&(1-U)\tau_{l+1}-2c1\omega\tilde{\omega}Q_{l+1}k_{lm}\left(il\bar{\mathcal{B}}^{'}-m\bar{\mathcal{B}}\right)-2c_1\omega\tilde{\omega}z_{l+1}\left[m\bar{\mathcal{B}}^{'}+i\bar{\mathcal{B}}\left((l+1)Q_{l+2}^2-(l+2)Q_{l+1}^2\right)\right]\nonumber\\
&&+c_1\omega s_{l+1}\left[\omega(1-\bar{\varepsilon})+2i\tilde{\omega}\bar{\mathcal{B}}\left((Q_{l+2}^2+Q_{l+1}^2)-1\right)\right]\label{dtau}\\
\zeta_{l+1}&=&-2i\omega\tilde{\omega}c_1\frac{l}{l+1}Q_{l+1}K_{lm}\label{orderz}\\
\tau_{l+1}(l+1)(l+2)&=&-2i\bar{\mathcal{B}}^{'}\omega\tilde{\omega}c_1{l}{(l+2)}Q_{l+1}k_{lm}+2{m}(l+2)\bar{\mathcal{B}}\omega\tilde{\omega}c_1Q_{l+1}k_{lm}\nonumber\\
&&+c_1\omega z_{l+1}\left\{\omega(l+1)(l+2)(1-\bar{\varepsilon})-2m\tilde{\omega}\bar{\mathcal{B}}^{'}-2i\tilde{\omega}\bar{\mathcal{B}}\left[m^2+Q_{l+2}^2(l+1)(l+4)+Q_{l+1}^2(l+2)(l-1)\right]\right\}\nonumber\\
&&-2c_1\omega\tilde{\omega}s_{l+1}\left\{m\bar{\mathcal{B}}^{'}-i\bar{\mathcal{B}}\left[Q_{l+2}^2(l+4)-Q_{l+1}^2(l-1)-1\right]\right\}\label{ordertau}\\
\eear
where
\bear
K_{lm}&=&-\frac{i\omega_0}{me}lQ_{l+1}\left[S_{l+1}+(l+2)Z_{l+1}\right]\\
k_{lm}&=&\frac{\omega_0}{m\eta}\left(m\bar{\mathcal{B}}-il\bar{\mathcal{B}}^{'}\right)Q_{l+1}\left[s_{l+1}+(l+2)z_{l+1}\right]
\eear
In the last two expressions we have used
\bear
e&=&(\omega-\omega_0)+3D_2\left\{\frac{2\omega}{l(l+1)}\left[lQ_{l+1}^2\right]-\omega_0\left[5Q_{l+1}^2-1\right]\right\}\label{defe}\\
\eta&=&(1-\bar{\varepsilon})(\omega_0-\omega_1)
\label{etadef}
\eear
It is easy to verify that these equations coincide with the results  of \cite{Saio1982} in the barotropic limit. In fact, in order to
facilitate this comparison we have used essentially the same notation as Saio. Thus,
we have defined the normalised mode frequency
\be
\omega =\sigma \left(\frac{R^3}{GM}\right)^{1/2}
\ee
and the following background quantities;
\bear
M_r &=& 4\pi \int_0^a \rho a^2 \, da \label{Mdef}\\
U &=& \frac{d \log M_r}{d \log a} \\
V &=& -\frac{d \log P}{d \log a} \equiv \frac{ga\rho}{P} \\
c_1 &=& \left(\frac{a}{R}\right)^3 \frac{M}{M_r}  \\
\tilde{\omega} &=& \Omega \left(\frac{R^3}{GM}\right)^{1/2} \\
X&=&-\frac{d\log \left[\rho x_\p(1-x_\p)\right]}{d\log a}\\
C&=&\frac{a}{x_\p(1-x_\p)}\frac{d x_\p}{da}\\
\epsilon &=& D_1(a) + D_2(a) P_2(\cos\theta) \label{epsdef}
\end{eqnarray}
where $P_2(\cos\theta)$ is the Legendre polynomial:
\be
P_2(\cos\theta) = (3\cos^2\theta -1)/2
\ee
We are also using the definition
\be
Q_l=\left[\frac{(l-m)(l+m)}{(2l+1)(2l-1)}\right]^{1/2}
\label{Qdef}
\ee
For later convenience, it is worth noting that
 this means that $Q_m=0$.

\subsection{Rotating $n=1$ polytropes}

We will now restrict ourselves to the case where the density profile of the equilibrium configuration is that of an $n=1$ polytrope.
This  greatly simplifies the analysis, as the background quantities we are interested in can be obtained in closed form.
Explicitly, we have
\be
p=K\rho^2
\ee
with
\be
K=\frac{2GR^2}{\pi}
\ee

Introducing the dimensionless variable $y=\pi a/R$ we then find that
\be
\rho=\frac{\pi M_0}{4R^3}\frac{\sin y}{y}
\ee
from which we can define
\be
M_r=\frac{M_0}{\pi}[\sin y -y\cos y]
\ee
Furthermore, we can calculate, following the classical work of \cite{Chandra}, the rotationally induced deformation
\be
D_1=\tilde{D}_1\tilde{\omega}^2=\frac{2}{\pi^2}\frac{M_0a\psi_1}{RM_r}\tilde{\omega}^2
\ee
and
\be
D_2=\tilde{D}_2\tilde{\omega}^2=-\frac{1}{9}\frac{M_0a\psi_2}{RM_r}\tilde{\omega}^2\label{didue}
\ee
where
\beq
\psi_1&=&1-\frac{\sin{y}}{y}\\
\psi_2&=&\frac{15}{y}\left[\left(\frac{3}{y^2}-1\right)\sin y-\frac{3}{y}\cos y\right]
\eeq
The mass of the star is thus, from equation (\ref{masse}), given by
\be
M=M_0\left[1+\frac{2}{\pi^2}\tilde{\omega}^2\left(\frac{\pi^2}{3}-1\right)\right]
\ee
In the following, when we quote the mass of the star we shall in fact be refering to the mass of the spherical star, $M_0$. Formally what we are doing is thus considering a sequence of stellar models with the same central density, but with a mass that varies with the rotation rate.
The difference in mass does not, however, enter the r-mode frequency correction to order $O(\Omega^3)$ and is thus irrelevant for the present discussion.

Finally, we will also need the sound speed, which simply follows from
\be
c_s^2 = 2K\rho
\ee

\subsection{Decoupling the degrees of freedom}\label{analytics}

Let us now examine the consequences of the ordering we  assumed at the beginning of the analysis. From equation (\ref{orderz}) we see that if $K_{lm}$ is of order unity then $\zeta_{lm}$ (i.e. the pressure perturbation) must be of order $\Omega^2$. Similarly, from equation (\ref{ordertau}) we see that, as all components of the difference displacement are of order $\Omega^2$, the variable $\tau_{lm}$  (i.e. $\delta\beta$) must be of order $\Omega^4$. It follows that in equation (\ref{dz}) the term involving $\tau_{lm}$ is of higher order than the others and can be neglected. The consequence of this is that the equations for the co-moving degree of freedom are  \underline{completely decoupled}. Hence,
they can be solved independently and then used as source terms for the counter-moving degrees of freedom.
This approach has obvious advantages compared to previous fully numerical calculations \citep{Lind3,yl2}.
After all, we can now solve for the r-mode throughout the star, imposing regularity at the centre of the star and the vanishing of the Lagrangian perturbation of the pressure, $\Delta p$, at the surface. As a second independent step, we solve the equations for the counter-moving degree of freedom. Adopting this strategy it becomes straightforward to account for,
for example, the temperature dependence of the superfluid gap and the associated variation of
the superfluid region in the star.

The solution for the comoving degree of freedom is particularly simple. Not only is it decoupled from the counter-moving degree of freedom, the equation for $\zeta_{lm}$ also decouples and takes the form:
\be
a\frac{d\zeta_{l+1}}{da}=(l+2-U)\zeta_{l+1}\label{zetan}
\ee
leading to the solution
\be
\frac{\zeta_{l+1}}{a}=2\frac{\omega\tilde{\omega}}{\sqrt{2l+3}}\frac{l}{l+1}\frac{M_0}{M_rR}\left(\frac{a}{R}\right)^{l+1}=B\frac{a^{l+1}}{M_r}
\ee
One can  use this solution to determine $S_{l+1}$ from equation (\ref{esse}). This leads to
\be
a\frac{dS_{l+1}}{da}+(l+4-V_g)S_{l+1}=-({V_g}+h)\zeta_{l+1}
\label{eqesse}\ee
where $V_g=V/\Gamma$ and (for the $l=m$ modes we are considering)
\be
h=\frac{1}{\omega_0^2}\frac{M(a)}{M}\left(\frac{R}{a}\right)^3\left[\frac{(l+1)(2l+3)e}{K\omega_0}+3\left(3D_2+a\frac{dD_2}{da}\right)\right]
\ee
with $e$ defined in equation (\ref{nome}).
To solve equation (\ref{eqesse}) let us first of all consider the solution to the homogeneous problem. This is readily obtained and takes the form
\be
S_{l+1}=\frac{C}{p^{1/\Gamma}a^{l+4}}
\ee
Note that this solution diverges both at the centre and  the surface of the star.
We next need to determine a particular solution to the problem. To do this let us write $h$ as
\be
h=\left(\frac{R}{a}\right)\frac{M(a)}{M}(\tilde{h}_c+\tilde{h}_a)
\ee
where
\beq
\tilde{h}_c&=&\frac{(l+1)(2l+3)\omega_2}{l\sigma_0^3}\\
\tilde{h}_a&=&\frac{3}{\sigma_0}\left[(2l+3)\tilde{D}_2+a\frac{d\tilde{D}_2}{da}\right]
\eeq
Inspired by the solution to the homogeneous problem we make the following ansatz for the particular solution to (\ref{eqesse});
\be
S_{l+1}=\frac{f(a)}{p^{1/\Gamma}a^{l+4}}
\ee
This leads to
\be
\frac{df}{da}=-(V_g+h)p^{1/\Gamma}a^{L+3}\zeta_{l+1}\label{df}
\ee
The first term is thus of the form
\be
V_gp^{1/\Gamma}a^{m+3}\zeta_{l+1}=BG\frac{a^{2l+4}}{\Gamma\sqrt{K}}
\ee
which is easily integrated. For the second part we need to be able to integrate, for the term proportional to $\tilde{h}_c$,
\be
\tilde{h}_cp^{1/\Gamma}a^{l+3}\zeta_{l+1}=B\sqrt{K}\tilde{h}_c\frac{R^3}{M}\rho a^{2l+2}
\ee
so that, for an $n=1$ polytrope we require
\be
\int_0^a\rho x^{2l+2} dx=\frac{\pi M}{4R^3}\left(\frac{R}{\pi}\right)^{2l+3}\int_0^{\pi a/R} y^{2l+1}\sin y dy=\frac{\pi M}{4R^3}\left(\frac{R}{\pi}\right)^{2l+3} \mathcal{I}_1(y)
\ee
For the remaining part, involving $\tilde{h}_a$, we require the integral
\be
\mathcal{I}_2=3\int\rho a^{2l+2}\left[(2l+3)\tilde{D}_2+a\frac{d\tilde{D}_2}{da}\right]
\ee
which, using the explicit form for $D_2$ in equation (\ref{didue}), gives
\be
\mathcal{I}_2=-\frac{5MR^{2l}}{4\pi^{2l+2}}[f_1(y)+f_2(y)]
\ee
where
\beq
f_1(y)&=&\frac{y^{2l}\sin y}{\sin y-y\cos y}[(3-y^2)\sin y-3y\cos y]\\
f_2(y)&=&\int_0^y x^{2l-1}[(3-x^2)\sin x - 3x\cos x] dx
\eeq

This analysis is perhaps a little bit messy, but the result is very useful.
Collecting the various results we can write the final solution to equation (\ref{df}) as
\be
f(y)=-\frac{BG}{\sqrt{K}}\left(\frac{R}{\pi}\right)^{2l+5}\left\{\frac{y^{2l+5}}{2(2l+5)}+\frac{\pi^2}{16}\frac{(l+1)^4(2l+3)}{l}\sigma_2\mathcal{I}_1(y)-\frac{5\pi^2(l+1)^2}{8}[f_1(y)+f_2(y)]\right\}
\ee
That is, we have an analytic solution to the problem.

As we now have the full solution to the problem we can impose boundary conditions.
First of all we require the solution to be regular at the centre of the star. This determines the constant $C=0$ in the homogeneous solution.
The remaining condition is that the Lagrangian variation of the pressure vanish at the surface of the star, which in terms of our variables means that
\be
\Delta p=\rho g a\sum [\zeta_{l+1}-S_{l+1}]Y_{l+1}^l\rightarrow 0\;\;\;\mbox{as}\;\;\;a\rightarrow 0
\ee
Given that  our equation of state is such that $\rho\rightarrow 0$ at the surface, we simply require $\zeta_{l+1}$ and $S_{l+1}$ to be regular. This
means that we must have
\be
f(\pi)=0
\ee
Conveniently, this  leads to an algebraic formula for the rotational frequency correction $\sigma_2$,
\be
\sigma_2=-\frac{16}{\pi^2}\frac{l}{(l+1)^4(2l+3)}\left[\frac{\pi^{2l+5}}{2(2l+5)}-\frac{5\pi^2(l+1)^2}{8}f_2(\pi)\right]\frac{1}{\mathcal{I}_2}
\ee
The determined frequency corrections for the first few values of $l$ are listed  in Table \ref{tab1}. These results are in good agreement with the numerical
results of \cite{LindOwen},  within a few percent for $l>2$. The greatest difference is  approximately $10\%$ for the $l=m=2$ mode.
The discrepancy should be entirely due to our use of the Cowling approximation. In fact for the $l=m=2$ mode our frequency agrees to within a few percent with that obtained with the numerical code of \cite{Andrea}.
Or results also agree well, with errors smaller than $1\%$, with the previous Cowling approximation result of \cite{review}.

\begin{table}
\begin{center}
\begin{tabular}{l | l l}
$l$&$\sigma_0$&$\sigma_2$\\
\hline
1&1.000&0.3033\\
2&0.666&0.4525\\
3&0.500&0.4429\\
4&0.400&0.4051\\
5&0.333&0.3659\\
6&0.286&0.3310\\
7&0.250&0.3010\\
8&0.222&0.2754\\
9&0.200&0.2535\\
10&0.182&0.2346
\end{tabular}
\end{center}
\caption{r-mode frequency corrections at second order in the slow-rotation approximation. The quantity $\sigma_2$ is defined by the relation $\sigma=\sigma_0\Omega+\sigma_2\Omega^3$ and, for the $l=m$ case that we are considering, $\sigma_0=2\Omega/(l+1)$.  These results are in good agreement with the results of Lindblom et al. (1999),  
within a few percent for $l>2$, the greatest difference is approximately $10\%$ for the $l=m=2$ mode. These differences are most likely due to our use of the Cowling approximation. In fact for the $l=m=2$ mode our frequency agrees to within a few percent with that obtained with the numerical code of Passamonti et al. (2008), in which the Cowling approximation is made, and to better than $1\%$ with the results of Andersson et al. (2001).}
\label{tab1}
\end{table}

\subsection{The superfluid degree of freedom}

At this point we have reached two interesting conclusions. First of
all, we have found that the rotational corrections for the r-mode can
be obtained in closed form for $n=1$ polytropes (in the Cowling
approximation). This analytic solution will undoubtedly help us
understand the nature of the r-modes and the effect of different
dissipation channels better. The second result concers the fact that
the superfluid, counter-moving, degree of freedom remains uncoupled
also at this order of approximation.  This is obvious since the
frequency correction to order $\Omega^3$ could be calculated without
determining the counter-moving eigenfunctions.  These are, in fact,
only needed if we want to calculate the frequency correction to order
$\Omega^5$. As we shall see in the following this is exactly the order
at which the mutual friction damping corrections appear explicitly in
the frequency.

To complete the r-mode solution at the current level of approximation
we need to use the analytic solution to the co-moving problem to
provide the source terms in equations (\ref{contro1}) and (\ref{fine})
for the counter-moving degrees of freedom.  When we consider this
problem it becomes apparent that the decoupling of the two degrees of
freedom is advantageous. In fact, it is now straightforward to study
neutron stars with different models of superfluidity. This is an
important improvement on previous work in this area, where it has
generally been assumed that the entire neutron star core is
superfluid. Our models will obviously still not be truly realistic,
but we can (at least) compare results for different predicted superfluid
energy gaps and varying neutron star core temperatures.

In order to determine to what extent the neutrons and protons are
superfluid/superconducting at a given core temperature (assumed to be
uniform), we will consider the phenomenological gap models discussed by
\citet{nuclphys}. It is straightforward to use these models since they
are given by analytic ``fits'' to the original results.
The only caveat is that the
fits are not valid at very low temperatures. One should also be
careful to use the analytic gap functions only in the density region
where they are relevant.

\begin{figure}
\centerline{\includegraphics[height=5.4cm,clip]{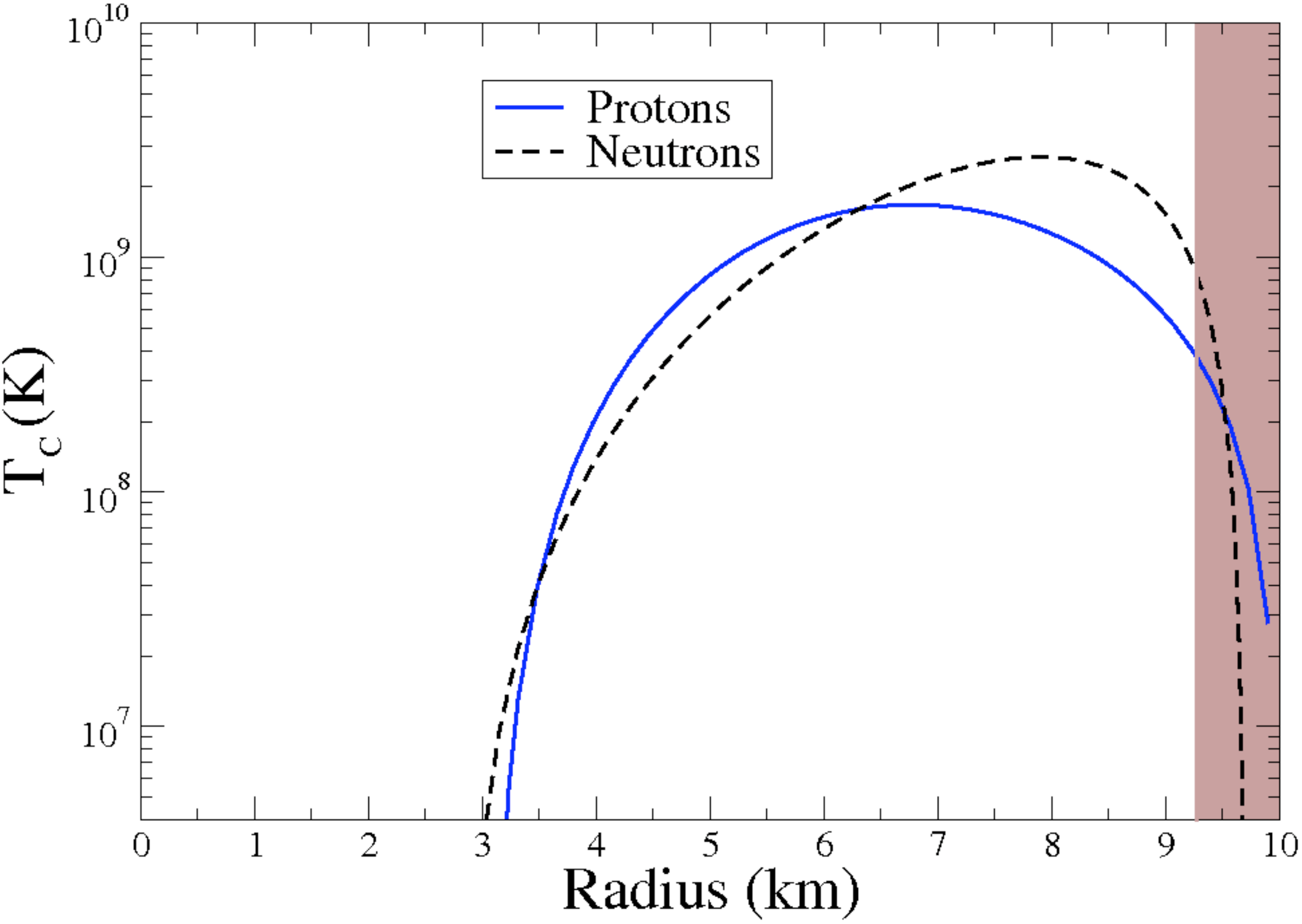}\includegraphics[height=5.4cm,clip]{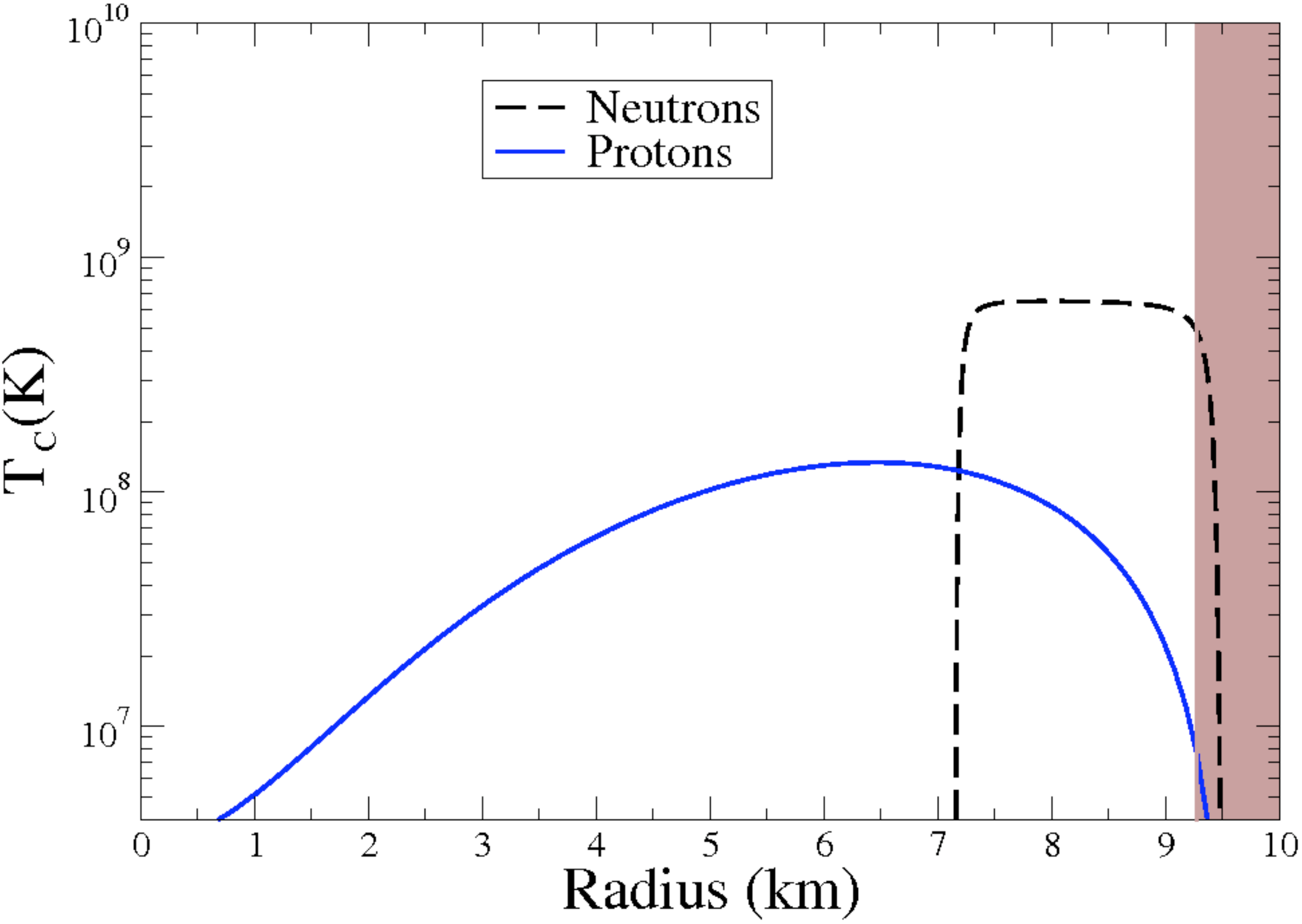}}
\caption{These figures show the critical temperatures for the onset of superfluidity of protons and neutrons as functions of radius in a stellar model with $M=1.4M_{\odot}$ and $R=10$~km. 
We consider two models representing ``strong'' (left) and  ``weak'' (right) superfluidity, respectively. 
For strong superfluidity we use models h and e of \citet{nuclphys}, while the weak superfluidity case corresponds to their models f and l. The shaded region indicates the crust, the dynamics of which 
is not accounted for in our r-mode calculation.}
\label{gaps}
\end{figure}

We will compare r-mode results for two different models, intended to
represent the range of possibilities.  These models correspond
to models [f,l] and [h,e] of \citet{nuclphys} and represent
 what we
will from now on refer to as ``strong'' and ``weak'' superfluidity,
respectively. For our polytropic neutron star model this leads to the
critical temperatures (as functions of the radius) shown in Figure
\ref{gaps}. The results correspond to a model with mass
$M=1.4M_{\odot}$ and radius $R=10$~km (in the non-rotating limit).
We take this as our canonical model. The
data in Figure~\ref{gaps} confirms the expectation that below a
critical temperature the superfluid region will always be a shell,
between to radii $R_\mathrm{min}(T)$ and $R_\mathrm{max}(T)$, where $T$ is the core temperature. In order to
simplify the analysis, we will only consider the core superfluids. This  makes sense
since we have not accounted for the elastic properties of the crust in the first place.
Hence, we consider only the $^3P_2$ neutron superfluid and the $^1S_0$ proton superconductor.
To model the fluid dynamics of this system, we make the usual assumption that protons and electrons
are locked electromagnetically. This leads to the system exhibiting two-fluid dynamics only in the region where the
neutrons are superfluid. When the neutrons are normal, scattering with the electrons will lock the neutron fluid to the charged
components. Given these assumptions the two-fluid region is such that $R_\mathrm{min}(T)$ corresponds to the inner
 $T=T_c$ point for the neutron gap. Meanwhile, $R_\mathrm{max}(T)$ corresponds to either the  outer
 $T=T_c$ point or the crust-core transition ($R_c$), whichever is the smallest. In our analysis we assume that the crust-core transition
takes place at a density of $\rho=1.6\times 10^{14}$ g/cm$^3$ \citep{Haensel}. For our canonical star this leads to $R_c\approx$9.32 km.

We need to impose boundary conditions on the two-fluid equations at $R_\mathrm{min}(T)$ and $R_\mathrm{max}(T)$. Following \cite{Lind3} and \cite{yl2} we
 take the pressure perturbation to be continuous at each  interface.
This ensures that it is consistent to use the analytic solution for the co-moving part of the solution throughout the star. With these interface conditions we
require
\be
\xi_- (R_\mathrm{min})=0\;\;\;\mbox{and}\;\;\;\xi_- \mathrm(R_\mathrm{max})=0
\ee
Naturally, the r-mode mutual friction damping time will now depend on the size of the superfluid shell, i.e. the temperature of the star.

We also want to be able to make a direct comparison with the results of  \cite{Lind3} and \cite{yl2}.
To do this we consider a model such that the entire core is superfluid, with a critical density  $\rho_s=2.8\times 10^{14}$ g/cm$^3$, corresponding to a transition radius $R_s\approx $\ 8.85 km in our  model.
We assume that  the pressure perturbation is continuous at this interface, i.e. impose
\be
\xi_- (R_s)=0
\ee
The remaining condition in this case  is regularity at the centre of the star.

\subsection{The mutual friction damping timescale}

We will use the standard energy integral approach to  assess the relative importance of different dissipation mechanisms
and their impact on the r-mode instability.  As discussed by, for example \citet{fmode}, this analysis is based on
a functional $E=E_k+E_p$,
where $E_k$  and $E_p$ represent the "kinetic" and "potential" energy, respectively.  In the non-dissipative case, the total energy is conserved.
For the  co-moving r-mode the problem simplifies considerably. The potential energy is higher order in rotation than the kinetic energy, so all we need is
\be
E \approx E_k=\frac{1}{2}\int \rho\left[|\delta v|^2+(1-\bar{\varepsilon})x_\p(1-x_\p)|\delta w|^2\right] dV
\label{Ek}
\ee
As discussed by \citet{fmode}, the damping timescale for  a mode $\tau=-1/\Im({\omega})$  follows from
\be
\frac{1}{\tau}=-\frac{1}{2E}\left(\frac{dE}{dt}\right)\label{dampo}
\ee
Focussing on the mutual friction dissipation, we find from equations (\ref{Eulav}) and (\ref{Euldi}) that
\be
\frac{dE}{dt}=-2\int\rho_\n\mathcal{B}\Omega[\delta^m_i-\hat{\Omega}^m\hat{\Omega}_i]\delta{w}^i\delta w^*_m dV
\label{dE}
\ee

At the current level of approximation, we cannot determine the mutual friction damping of the classical r-mode
 directly from the imaginary part of the mode frequency. This would require calculating the frequency  up to order $O(\Omega^5)$, not
$O(\Omega^3)$ as we have done here.
This is easily seen if we note that, as the mode is purely co-moving to first order, $\delta v\approx O(\Omega)$ and $\delta w\approx O(\Omega^3)$.
Thus, the energy scales as
\begin{displaymath}
E\approx (\delta v)^2\approx O(\Omega^2)\;\;\;\mbox{while}\;\;\;\frac{dE}{dt}\approx \Omega(\delta w)^2\approx O(\Omega^7)
\end{displaymath}
and, from equation (\ref{dampo}), we see that $\Im({\omega})\approx O(\Omega^5)$.
This means that we, in fact, have to resort to the energy integral estimate. Since the evaluation, to leading order,
of the integrals in (\ref{Ek}) and (\ref{dE}) only requires the eigenfunctions up to order $O(\Omega^2)$ it can be performed within the present scheme.
Due to the expected scaling we will focus on the quantity $\tau_0$ defined by the relation
\be
\frac{1}{\tau}=\frac{1}{\tau_0}\left(\frac{\Omega}{\sqrt{\pi G \bar{\rho}}}\right)^5
\ee
where $\bar{\rho}$ is the mean density of the star.

\begin{figure}
\centerline{\includegraphics[height=8cm,clip]{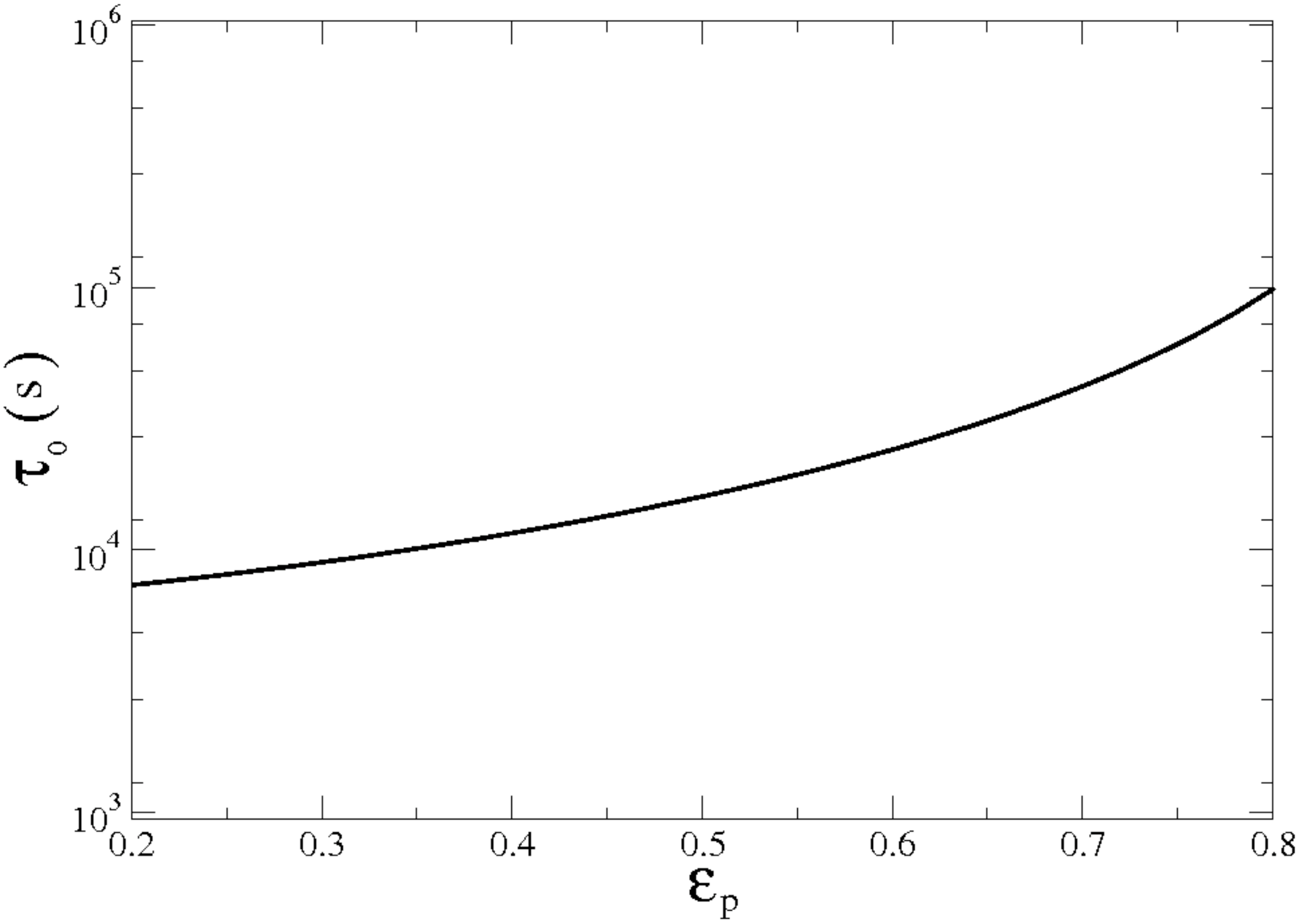}}
\caption{This figure shows the r-mode damping timescale, $\tau_0$, for a range of values of the entrainment parameter, $\varepsilon_\p$. 
In order to facilitate a direct comparison with the results of  \citet{Lind3} and \citet{yl2}, we consider a model where the entire 
core is superfluid. The transition density is taken to be  $\rho_s=2.8\times 10^{14}$ g/cm$^3$. The stellar model has mass $M=1.4M_{\odot}$ and  radius $R=12.533 $~km. 
We also assume that the mutual friction is due to electron scattering on vortex array, which gives $\mathcal{B}^{'}\approx 0$ while $\mathcal{B}$ is obtained from equation (\ref{trevmf}).
The absence of ``resonances'', associating particular values of the entrainment with very short damping times, is notable in our results.}
\label{damp}
\end{figure}

Having established the strategy, let us
first of all consider the model where the superfluid extends all the way from the centre of the star up to $R_s$. In order to
compare our results directly to  \cite{Lind3} and \cite{yl2}
we take the mass and radius of the star to be $M=1.4M_{\odot}$ and  $R=12.533$ km. We also assume that the main cause of
mutual friction is electron scattering off the vortex array. This assumption puts us firmly in the "weak" drag regime
where we can neglect the  $\mathcal{B}^{'}$ coefficient. Meanwhile, for $\mathcal{B}$ we have the result of \cite{trev};
\be
\mathcal{B}=4\times 10^{-4}\left(\frac{m_\p- m^{*}_\p}{m_\p}\right)^2\left(\frac{m_\p}{m_\p^{*}}\right)^{1/2}\left(\frac{x_\p}{0.05}\right)^{7/6}\left(\frac{\rho}{10^{14}\ \mathrm{g/cm}^3}\right)^{1/6}
\label{trevmf}
\ee
Results for the damping timescale $\tau_0$ due to mutual friction, with varying $\varepsilon_\p$, are presented in Figure~\ref{damp}.
These results  confirm  the main conclusion of \cite{Lind3} and \cite{yl2}. Mutual friction due to electron scattering off the vortex array
is too weak to affect the r-mode instability significantly. There is, however, one important difference between our results and the previous studies.
We do not find any value of $\varepsilon_\p$ for which the damping timescale becomes  short, cf. the resonances discussed by \cite{Lind3} and \cite{yl2}.
The absence of these resonances is most likely due to the fact that, following  \citet{fmode}, we have imposed that the mode is
purely axial to leading order. That is, we neglect higher order terms that would couple the counter-moving motion back to the co-moving motion.
In effect, we have a priori ruled out the  resonant behaviour found by \cite{Lind3}. Moreover, we cannot have  avoided crossings
with the superfluid inertial modes as discussed by \cite{yl2}. Once we assume that the r-mode is purely axial to leading order, our analysis becomes
completely oblivious to the fact that other mode solutions to  the general problem may exist.

\begin{figure}
\centerline{\includegraphics[height=8cm,clip]{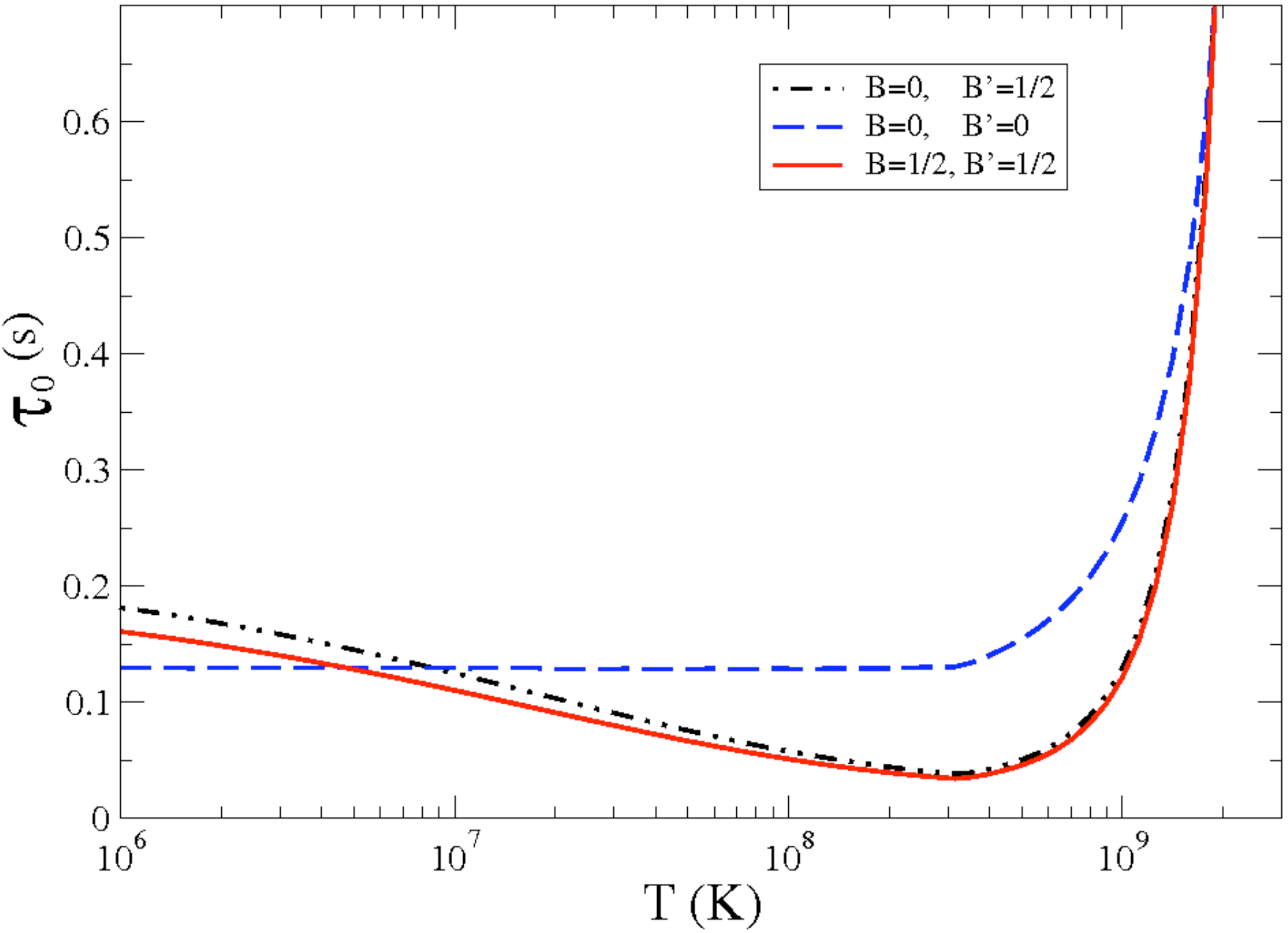}}
\caption{This figure shows the r-mode damping timescale, $\tau_0$, calculated for $\varepsilon_\p$=0.6 and $\mathcal{R}=1$ for the strong superfluidity model.  
The results illustrate the importance of using the dissipative mode-solution in the evaluation of the energy integrals.
The dashed curve shows the timescale obtained by integrating the undamped eigenfunctions. The other  two curves show the effect of introducing first $\mathcal{B}'$ (dash-dot) and then also
$\mathcal{B}$ (solid) in the calculation.}
\label{damp2}
\end{figure}

Another important difference between our analysis and the work of \cite{Lind3} and \cite{yl2} is that we explicitly keep the mutual friction terms in the equations of motion.
The damping timescale we calculate is obtained by using the full $O(\Omega^3)$ solution for the problem in the integral in equation (\ref{dampo}),
 rather than  integrating the solutions to the inviscid problem. In the weak drag limit, e.g. when we consider (\ref{trevmf}), this does not affect the results at all. 
However, by keeping the mutual friction force in the
 perturbed equations of motion we are no longer restricted to the weak drag limit. We can consider the entire range $0\le \mathcal{R} \le \infty$. The capacity to consider the strong drag regime
 may, in fact, be quite important. Dynamics in the strong drag regime has only recently been considered, see e.g. \cite{letter}, and the results show that the problem has interesting features. Hence, we will
 consider the r-mode problem outside the range $\mathcal{R}\ll 1$. Motivation for studying the problem
 for $\mathcal{R}\gg 1$ comes from the possibility that the interaction between fluxtubes and neutron vortices
 may be efficient \citep{strong1,strong2,strong3}. The problem would also be in this strong drag regime if a fluxtube cluster is associated with each neutron vortex \citep{strong4} or in the presence of strong vortex pinning \citep{strong5}.

Because of the ``symmetric'' dependence of the mutual friction coefficient $\mathcal{B}$ on the drag
$\mathcal{R}$, one may expect the damping timescales to be similar for values $\mathcal{R}$ and $1/\mathcal{R}$.
The damping should be most effective for $\mathcal{R}\approx 1$. The results in Figure~\ref{damp2}
illustrate the importance of
keeping the mutual friction force terms in the perturbed equations of motion in this efficient friction regime.
In the figure we compare the mutual friction damping timescale determined from the energy integral using eigenfunctions where we either keep both the $\mathcal{B}$ and the $\mathcal{B}'$ terms or (artifically) set one or both of them to zero. The evidence is clear. For $\mathcal{R}\approx 1$  the inviscid solution does no longer lead to a good approximation.

\section{The r-mode instability window for strong mutual friction}

It is interesting to consider the case $\mathcal{R}\approx 1$ further, even though it is somewhat extreme (and possibly only of academic interest). After all, it leads to the strongest possible mutual friction damping.
Moreover, because it corresponds to $\mathcal{B} \approx \mathcal{B}' \approx 1/2$ it provides insight into the
dynamics of problems where the two terms in the mutual friction force are of similar importance.

Let us consider the effect that a strong mutual friction would have on the r-mode instability.
To do this we need to account for the various viscous processes which, at a given temperature, act to damp the gravitational-wave driven instability.
First of all it is easy to see that the r-modes will only be unstable in a certain temperature range. At temperatures below $T\approx 10^5$ K shear viscosity will always suppress the instability, while for temperatures above $T\approx 10^{10}$ K (at which no part of the star is expected to be superfluid) bulk viscosity will prevent the mode from becoming unstable.
Our aim is to establish to what extent the mutual friction can further restrict the range in which the mode is unstable.

The r-mode instability window is usually illustrated by the critical rotation period above which the mode is unstable
as a function of temperature. The relevant critical rotation rate is obtained by solving for the roots of
\be
\frac{1}{\tau_\mathrm{gw}}+\frac{1}{\tau_\mathrm{sv}}+\frac{1}{\tau_\mathrm{bv}}+\frac{1}{\tau_\mathrm{mf}}=0
\ee
where $\tau_\mathrm{gw}$ is the growth timescale for the instability due to gravitational-wave emission. For an $n=1$ polytrope it is approximated by \citep{review}
\be
\tau_\mathrm{gw}\approx -47\left(\frac{M}{1.4M_{\odot}}\right)^{-1}\left(\frac{R}{10\mbox{ km}}\right)^{-2l}\left(\frac{P}{1\ \mathrm{ms}}\right)^{2l+2} \mbox{s}
\ee
Even though this estimate was obtained for a single fluid star, it will remain a good approximation for the
ordinary r-modes of a superfluid star. This is obvious since the gravitational-wave emission is entirely due to the co-moving degree of freedom \citep{fmode}. Meanwhile,
 $\tau_\mathrm{bv}$ represents the damping time due to bulk viscosity. The bulk viscosity will be suppressed in superfluid regions. Hence, it is only active in regions where the star is  not superfluid.
 In general, this leads to a complex problem. However, in the case of npe-matter where only the modified URCA process
 is relevant the bulk viscosity damping is not important at low temperatures. Hence, we essentially only need the bulk viscosity at temperatures above the superfluid transition. That is, we use \citep{review}
\be
\tau_\mathrm{bv}=2.7\times 10^{11} \left(\frac{M}{1.4M_{\odot}}\right)\left(\frac{R}{10\mbox{ km}}\right)^{-1}\left(\frac{P}{1\ \mathrm{ms}}\right)^{2}\left(\frac{T}{10^9 \mathrm{K}}\right)^{-6} \mbox{s}
\ee
It should be noted that the role of bulk viscosity will be different for hyperons and deconfined quarks because of a different scaling with temperature, see e.g. \citet{Owen2006} for a recent discussion.

The damping timescale, $\tau_\mathrm{mf}$, for mutual friction damping, and $\tau_\mathrm{sv}$, the timescale for damping due to shear viscosity require a more detailed discussion.
In the case of mutual friction we are interested in the temperature dependence of the damping timescale. To investigate this we consider the realistic gap models described in the previous section which, as they predict how the extent of the superfluid region varies with temperature, allow us to predict the temperature dependence of $\tau_\mathrm{mf}$ itself.
Meanwhile, the shear viscosity damping also depends on which layers of the star are superfluid, as the dominant effect will be due to different processes in superfluid and normal fluid regions.
In the region where the neutrons are not superfluid (the normal fluid region) we assume that the main contribution is due to neutron-neutron scattering. This leads to a viscosity coefficient \citep{review}
\be
\eta_\mathrm{nn}=2\times 10^{18} \left(\frac{\rho}{10^{15} \mbox{g}/\mbox{cm}^3}\right)^{9/4}\left(\frac{T}{10^9 \mathrm{K}}\right)^{-2}\mbox{g/cm s}
\ee
If only the neutrons are superfluid then the dominant process will be electron-proton scattering which leads to the coefficient  (\cite{nuclphys})
\be
\eta_\mathrm{ep}=1.8\times 10^{16} \left(\frac{x_\p}{0.01}\right)^{13/6}\left(\frac{\rho}{10^{15} \mbox{g}/\mbox{cm}^3}\right)^{13/6}\left(\frac{T}{10^9 \mathrm{K}}\right)^{-2}\mbox{g/cm s}
\ee
If, on the other hand, \underline{both} neutrons and protons are superfluid the dominant process is electron-electron scattering \citep{nuclphys},  which leads to the coefficient \citep{review}
\be
\eta_\mathrm{ee}=6\times 10^{18} \left(\frac{\rho}{10^{15} \mbox{g}/\mbox{cm}^3}\right)^{2}\left(\frac{T}{10^9 \mathrm{K}}\right)^{-2}\mbox{g/cm s}
\ee
Once we have determined the dominant viscosity agent in each region of the star
we can compute the damping timescale from the energy integral
\be
\left[\frac{dE}{dt}\right]_\mathrm{sv}=-2\int \eta \delta\sigma^{ab}\delta\sigma_{ab}^{*} dV
\ee
where the shear $\delta\sigma_{ab}$ is defined as
\be
\delta\sigma_{ab}=\frac{i\omega}{2}\left(\nabla_a\xi^{+}_b-\nabla_b\xi^{+}_a-2g_{ab}\nabla_c\xi^c_{+}\right)
\ee

\begin{figure}
\centerline{\includegraphics[height=6.5cm,clip]{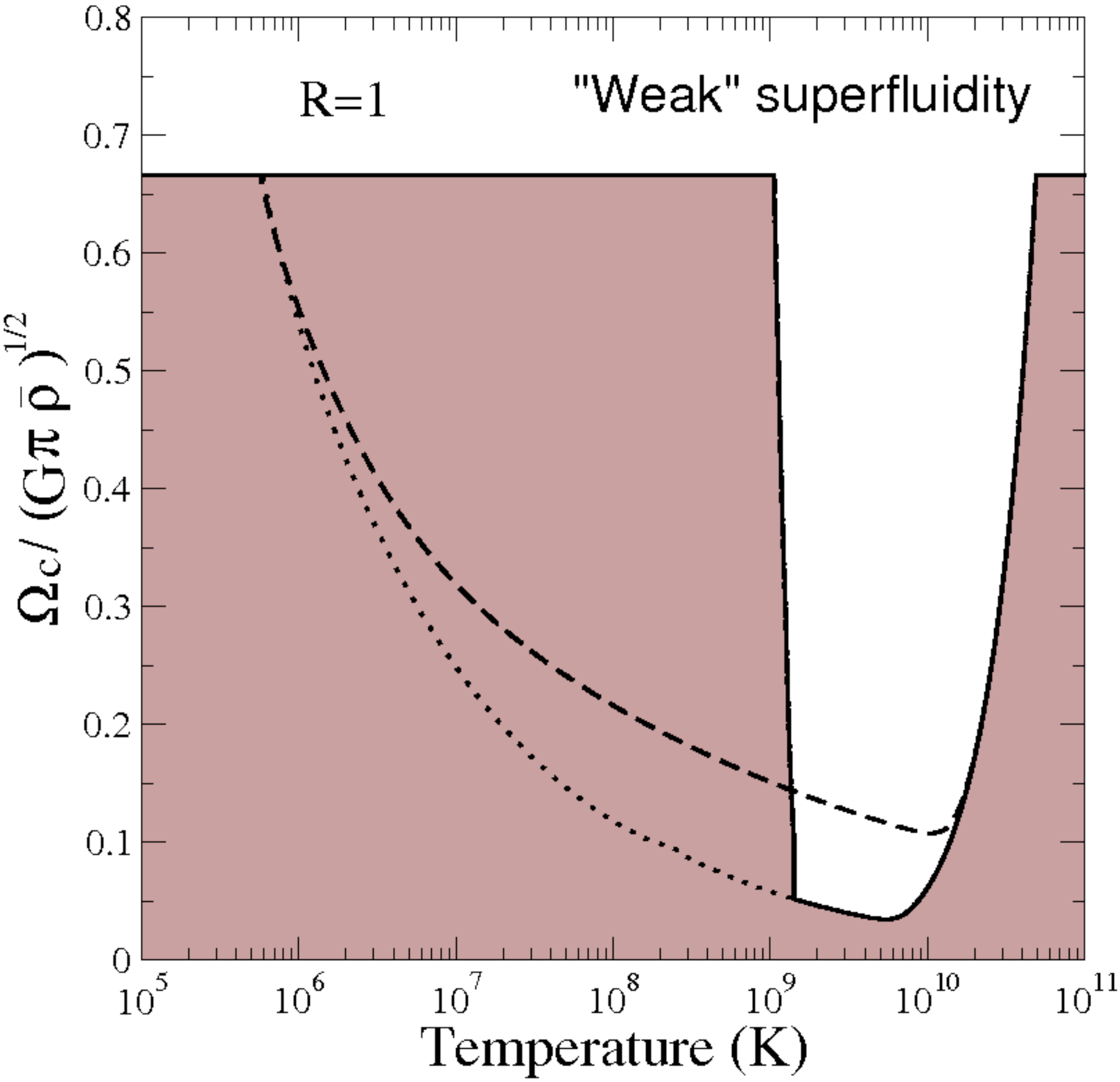}\includegraphics[height=6.5cm,clip]{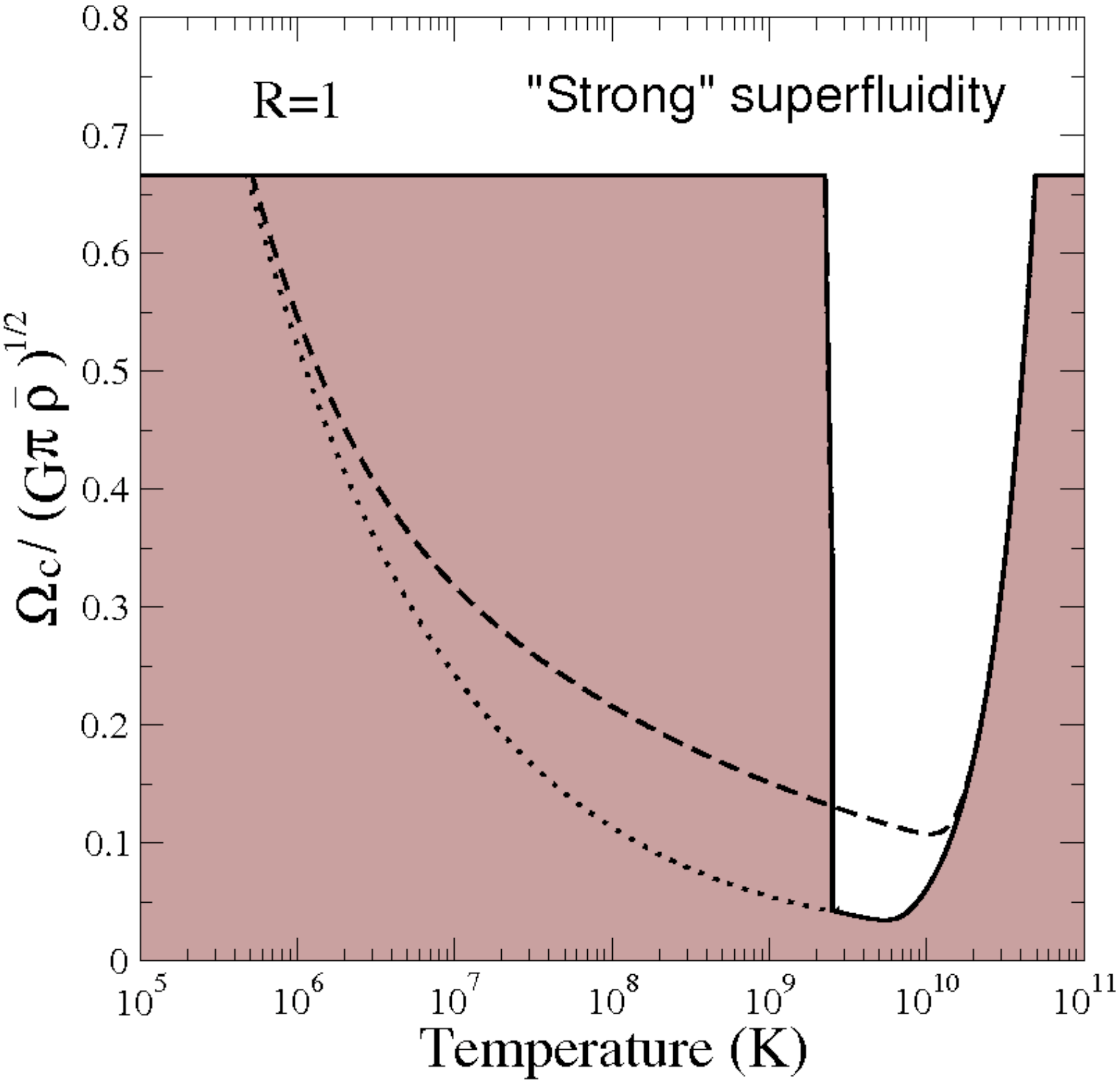}}
\caption{The r-mode instability window for  $\varepsilon_\p=0.6$, calculated as a function of core temperature $T$ and for $\mathcal{R}=1$ in the ``weak'' (left) and ``strong'' (right) superfluidity cases. 
We consider a star with $M=1.4M_{\odot}$ and $R=10$ km. The dotted line indicates the shape the instability window has if we ignore mutual friction, while the dashed line indicates the effect that the 
Ekman layer at the base of the crust would have on the instability region. The results show that, when $\mathcal{R}=1$, the mutual friction is strong enough to suppress the 
r-mode instability completely as soon as the core becomes superfluid.}
\label{swr1}
\end{figure}

\begin{figure}
\centerline{\includegraphics[height=5.5cm,clip]{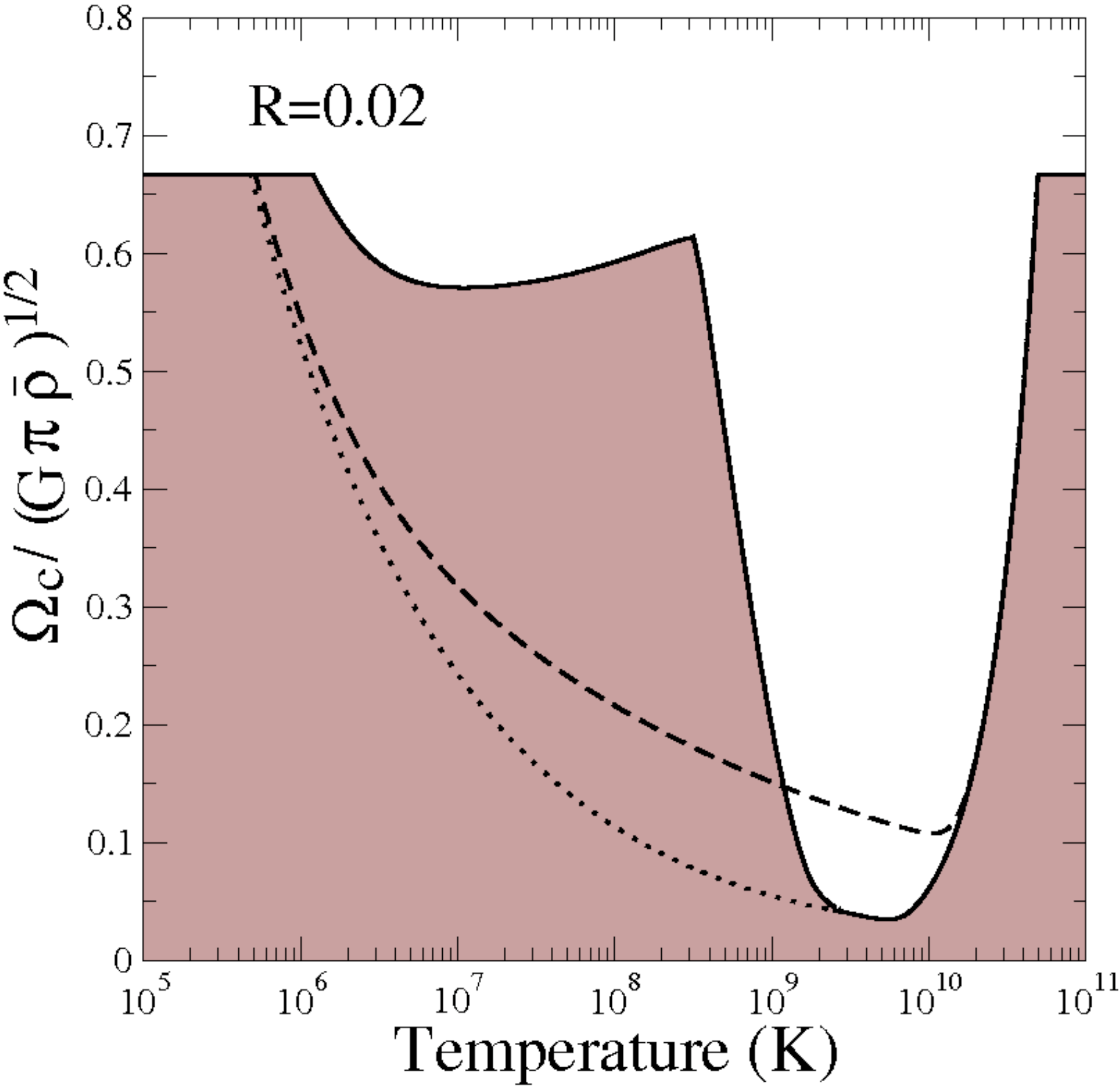}\includegraphics[height=5.5cm,clip]{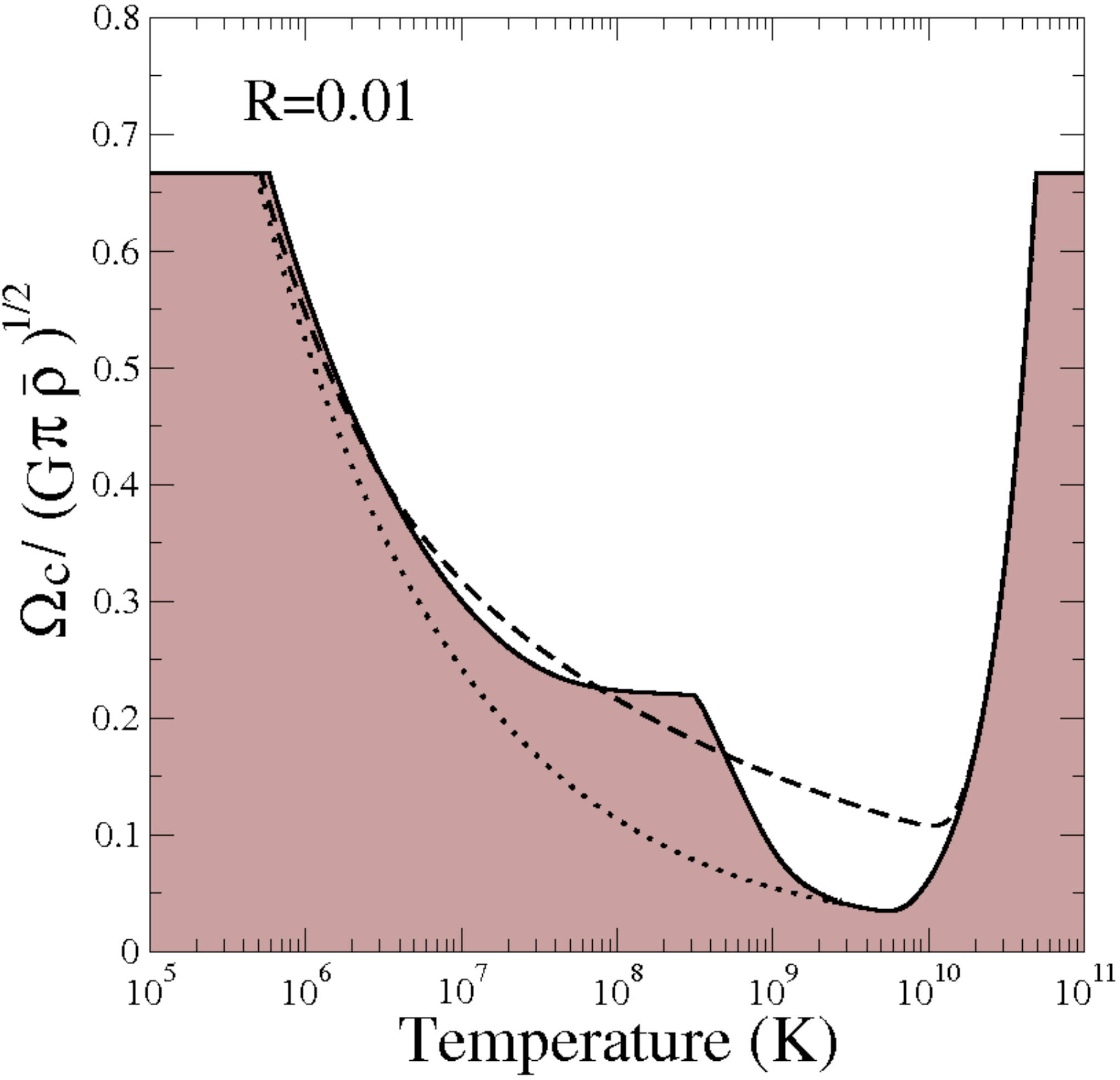}\includegraphics[height=5.5cm,clip]{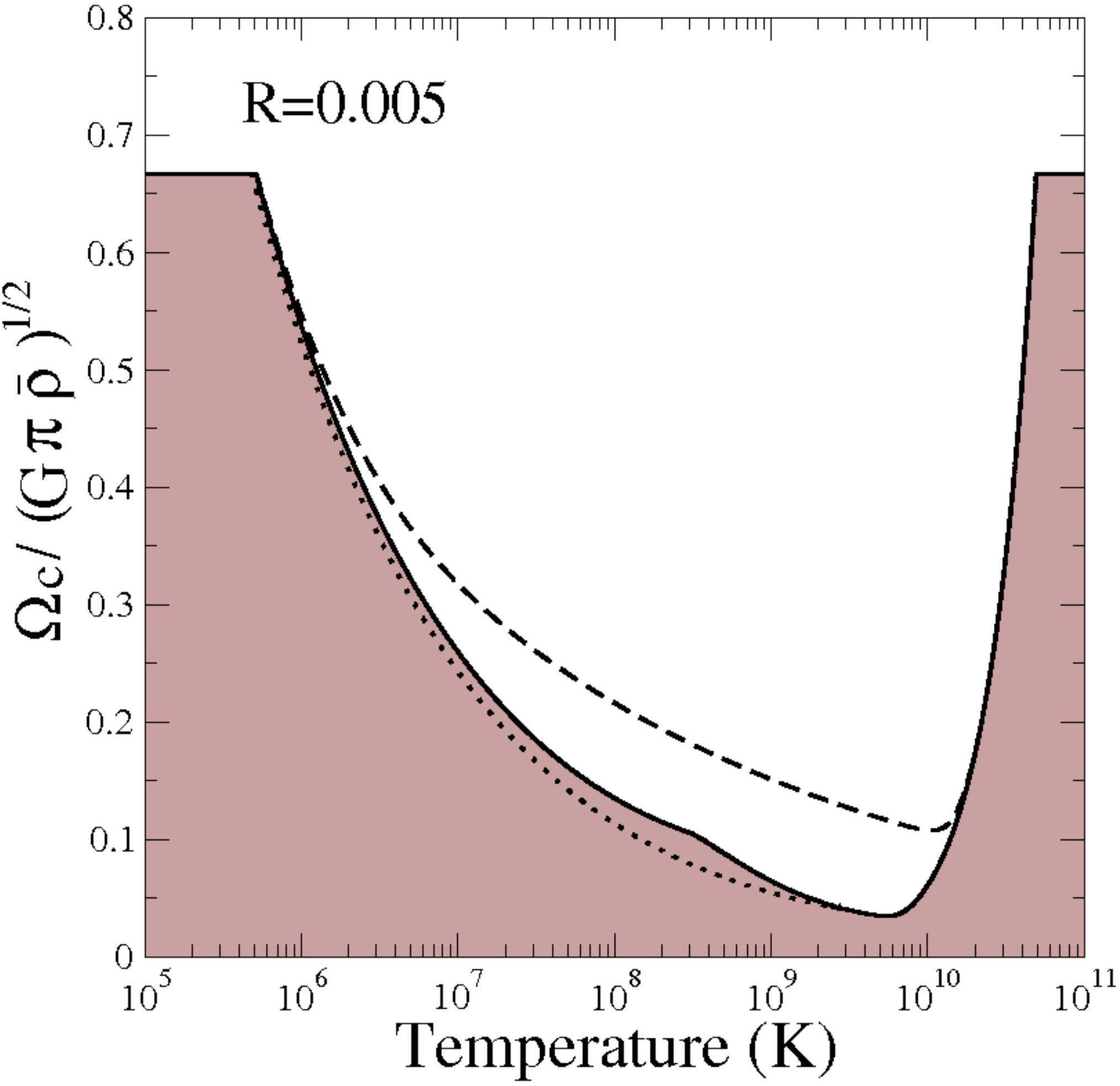}}
\centerline{\includegraphics[height=5.5cm,clip]{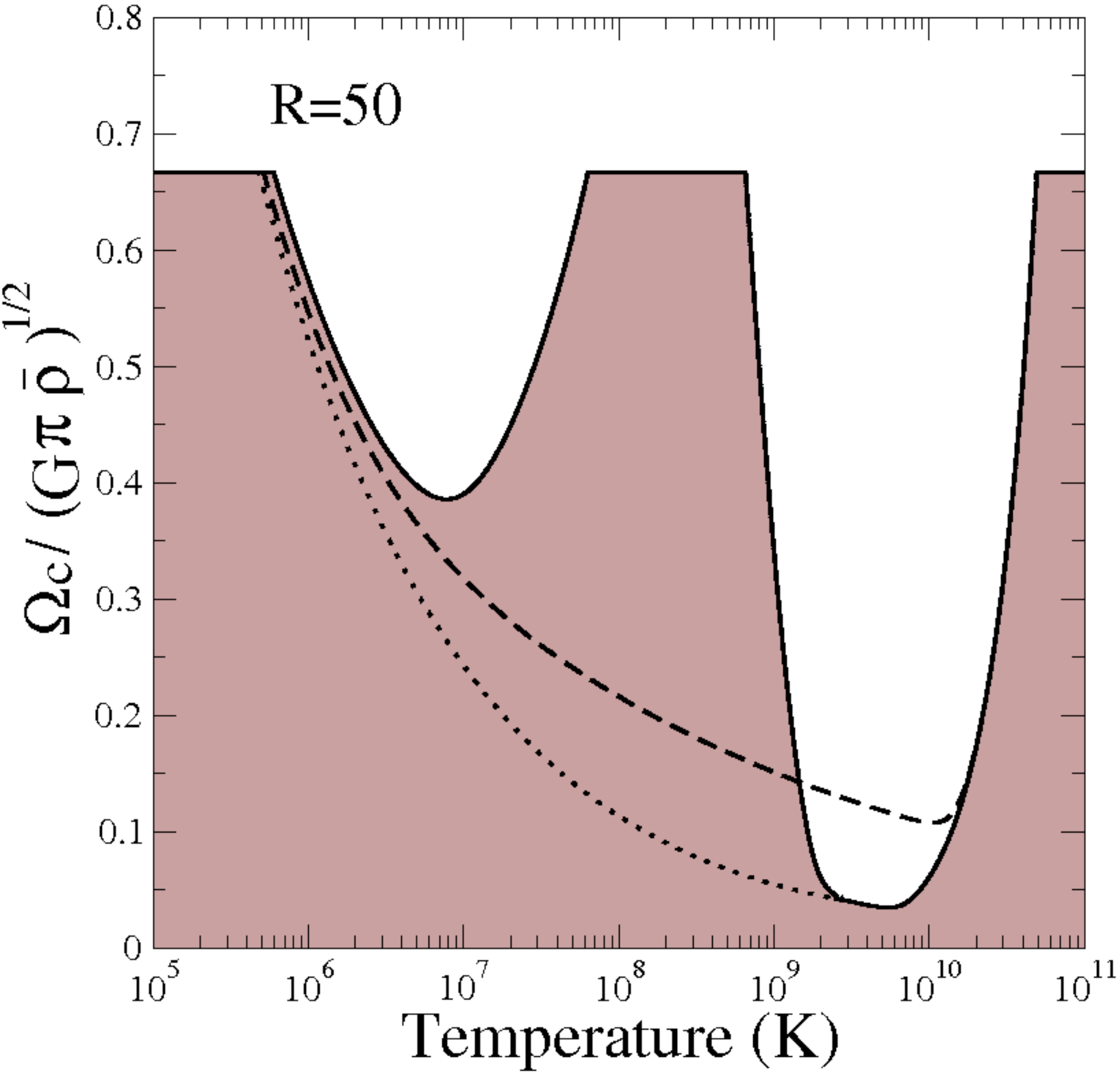}\includegraphics[height=5.5cm,clip]{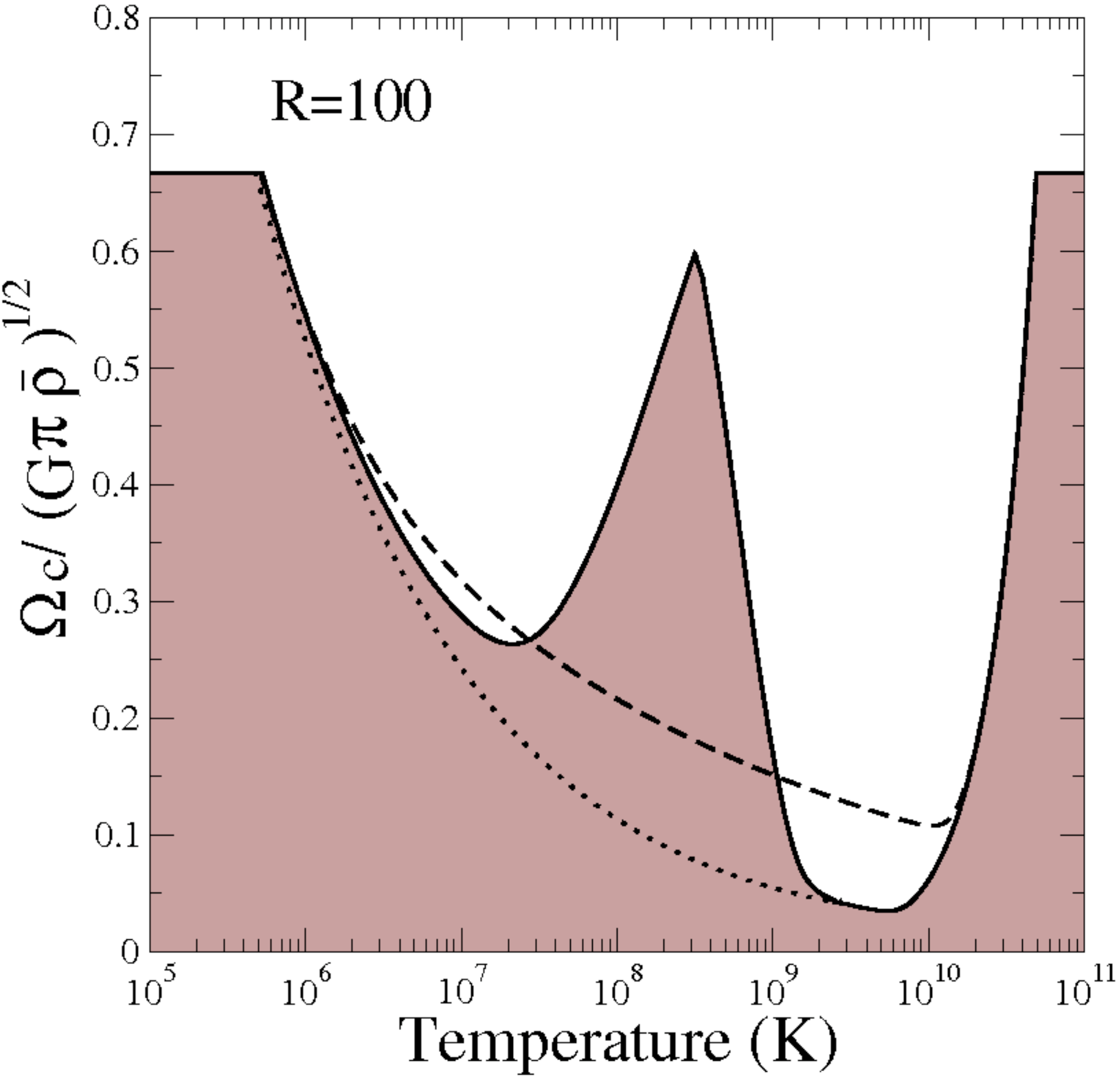}\includegraphics[height=5.5cm,clip]{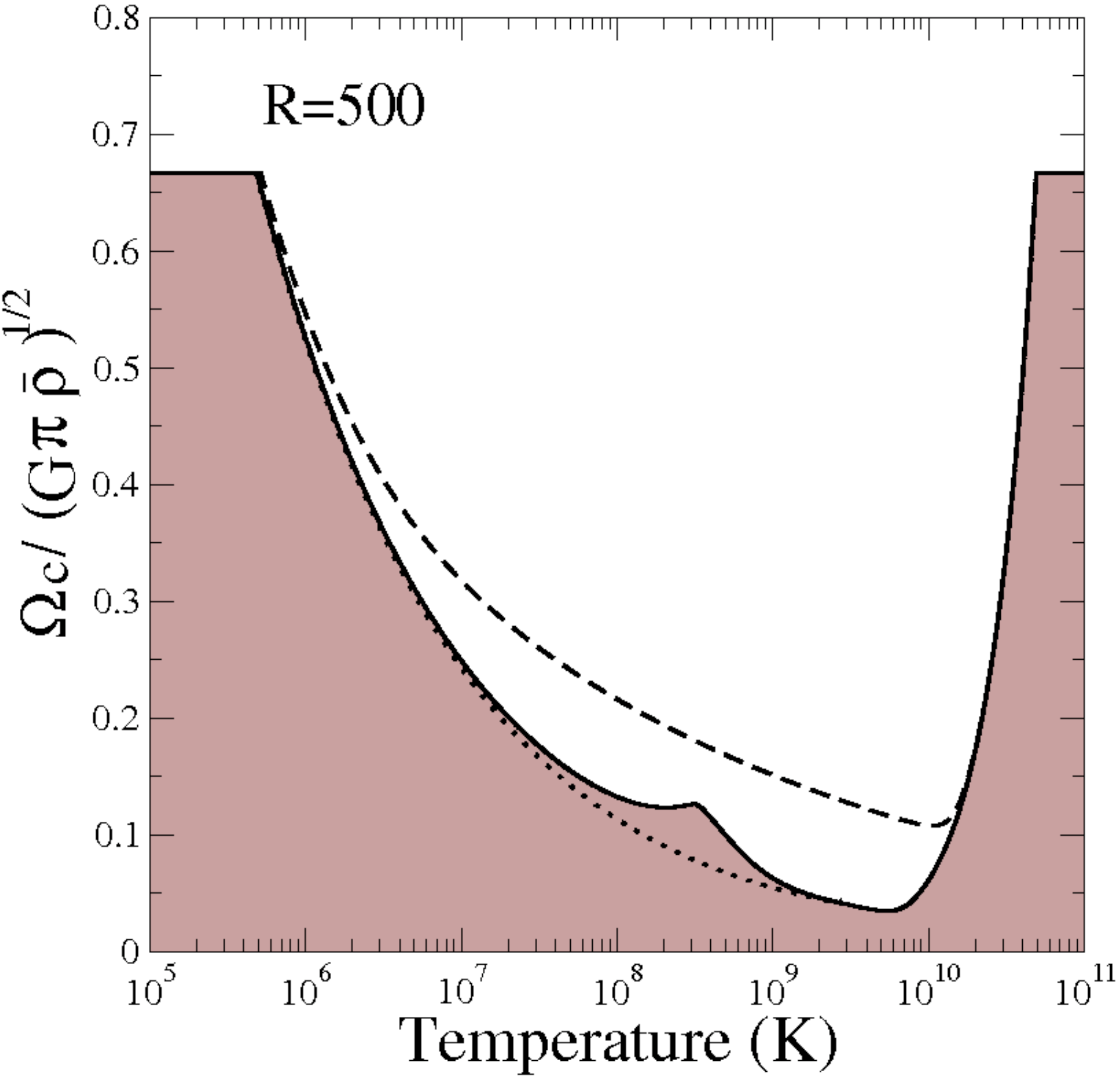}}
\caption{The r-mode instability window for  $\varepsilon_\p=0.6$, calculated as a function of core temperature $T$ and for a range of drag parameters $\mathcal{R}$. 
We consider a star with $M=1.4M_{\odot}$ and $R=10$ km and ``strong'' superfluidity.  The dotted line indicates the shape the instability window has if we ignore mutual friction, while the dashed line indicates the effect that the 
Ekman layer at the base of the crust would have on the instability region. The results show that, for values of the drag parameter in the range $0.005 < \mathcal{R} < 500$, mutual friction may have a significant effect on the r-mode instability. It is particularly interesting to note the local minimum that may be present at low temperatures. This feature arises due to the effect shown in figure~\ref{damp2}.}
\label{sw}
\end{figure}

\begin{figure}
\centerline{\includegraphics[height=5.5cm,clip]{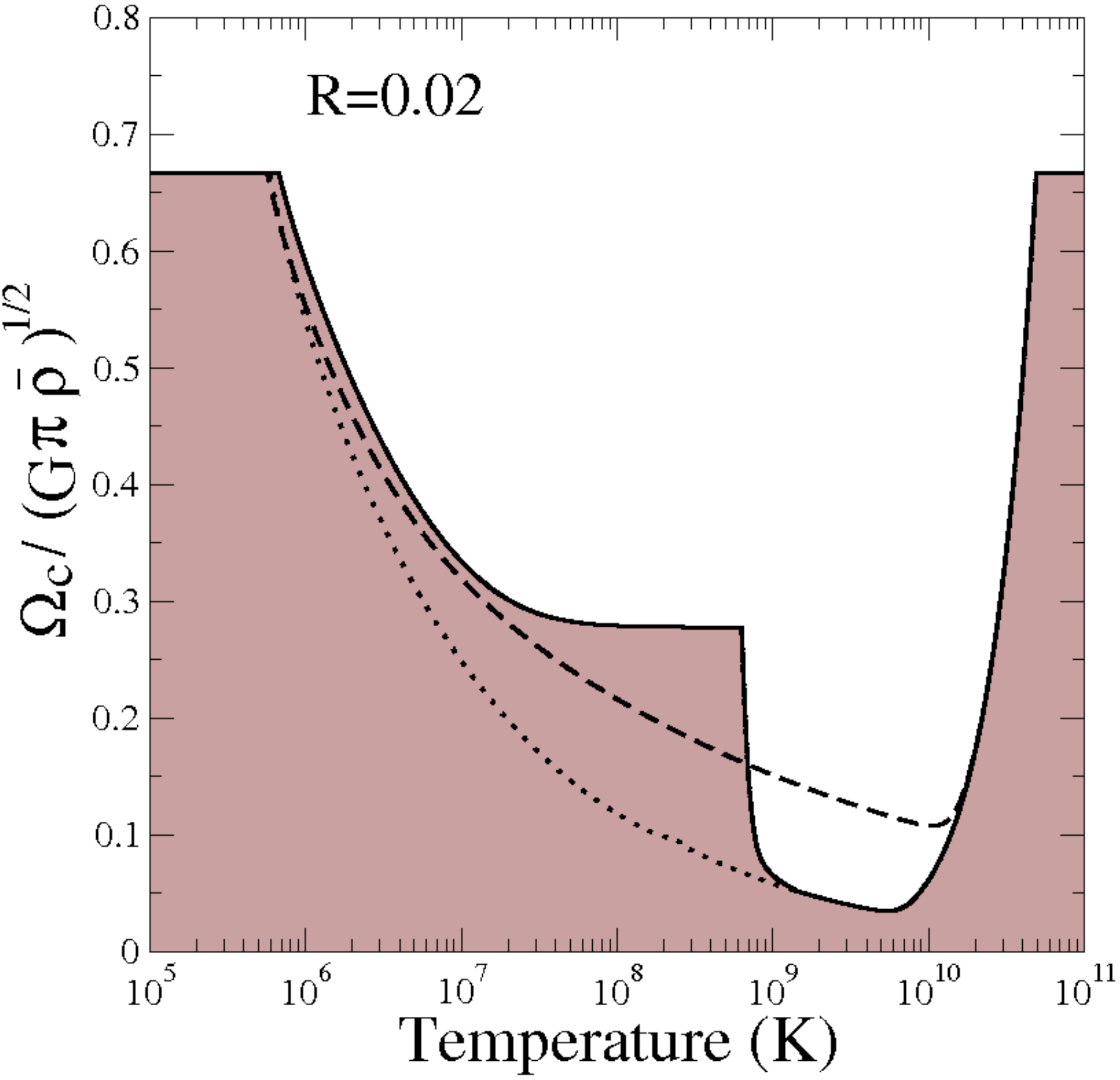}\includegraphics[height=5.5cm,clip]{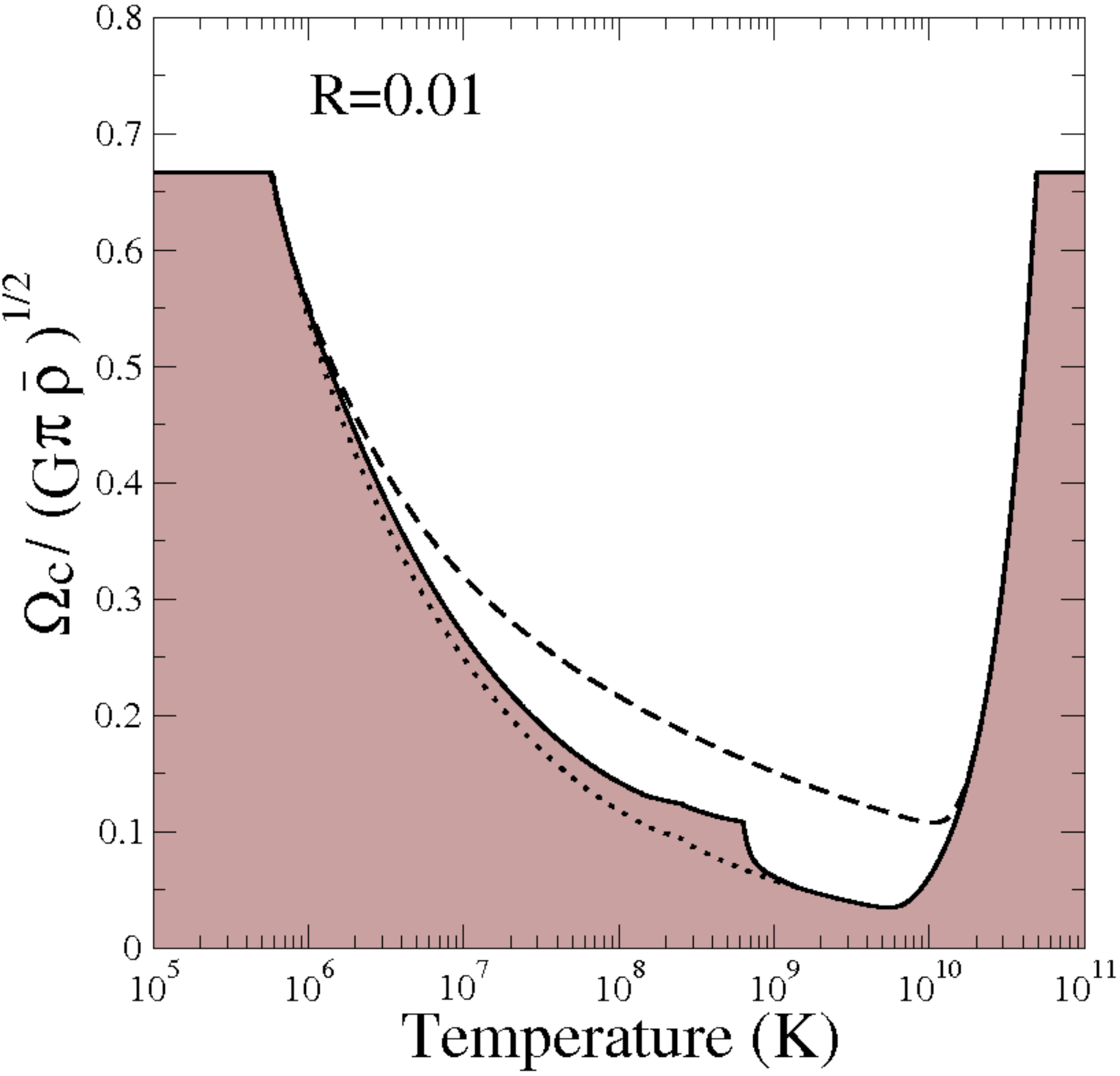}\includegraphics[height=5.5cm,clip]{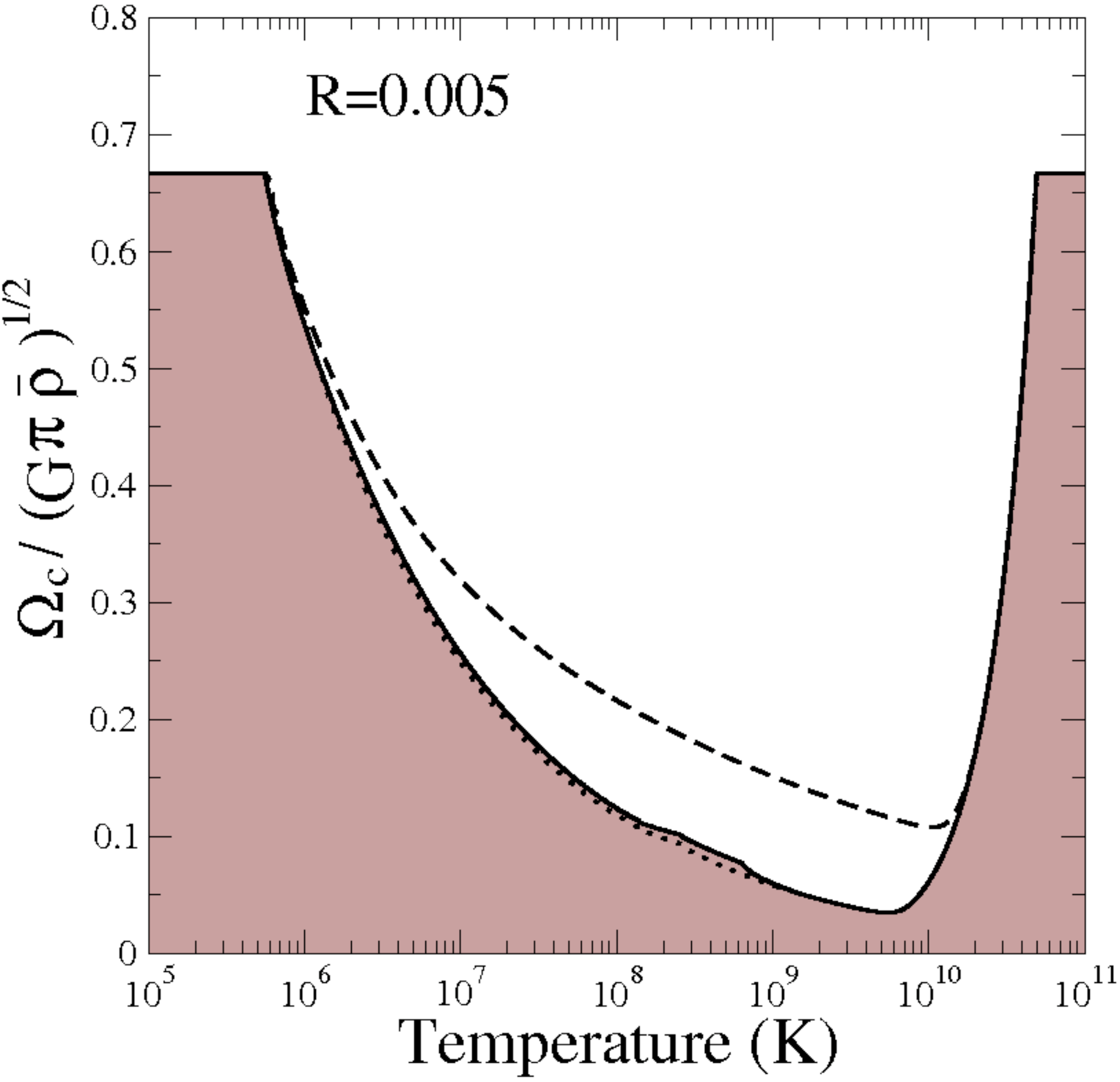}}
\centerline{\includegraphics[height=5.5cm,clip]{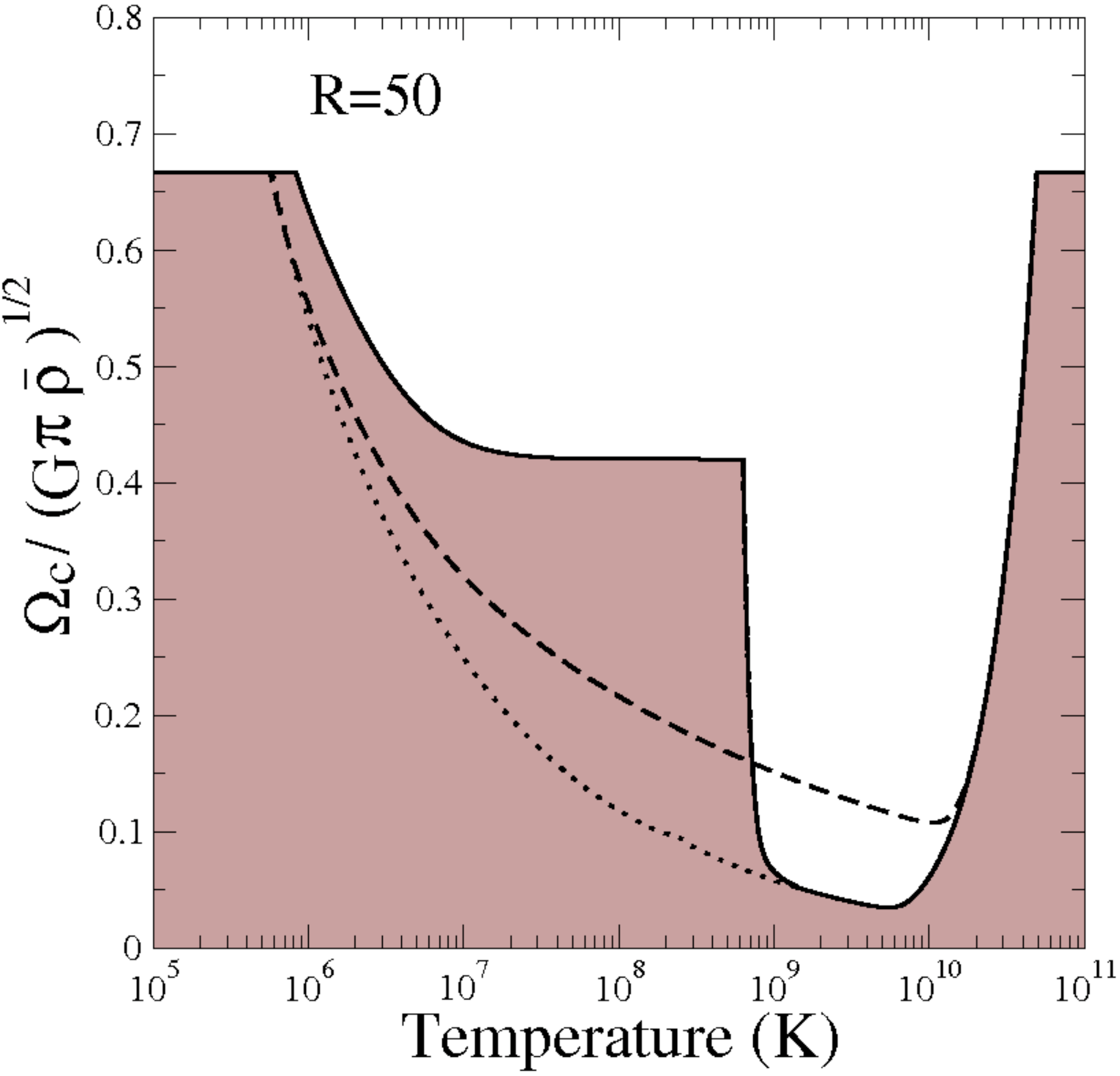}\includegraphics[height=5.5cm,clip]{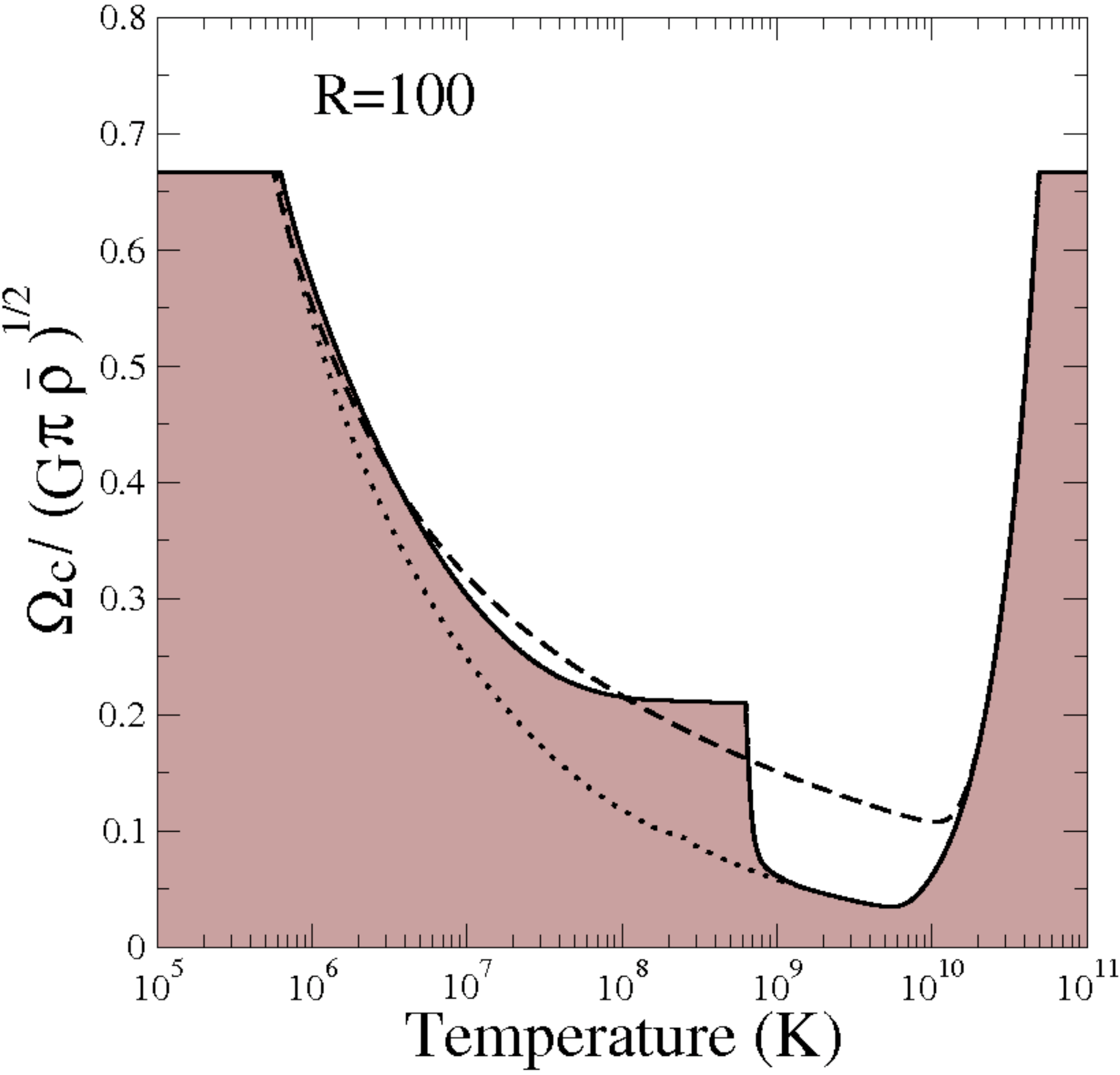}\includegraphics[height=5.5cm,clip]{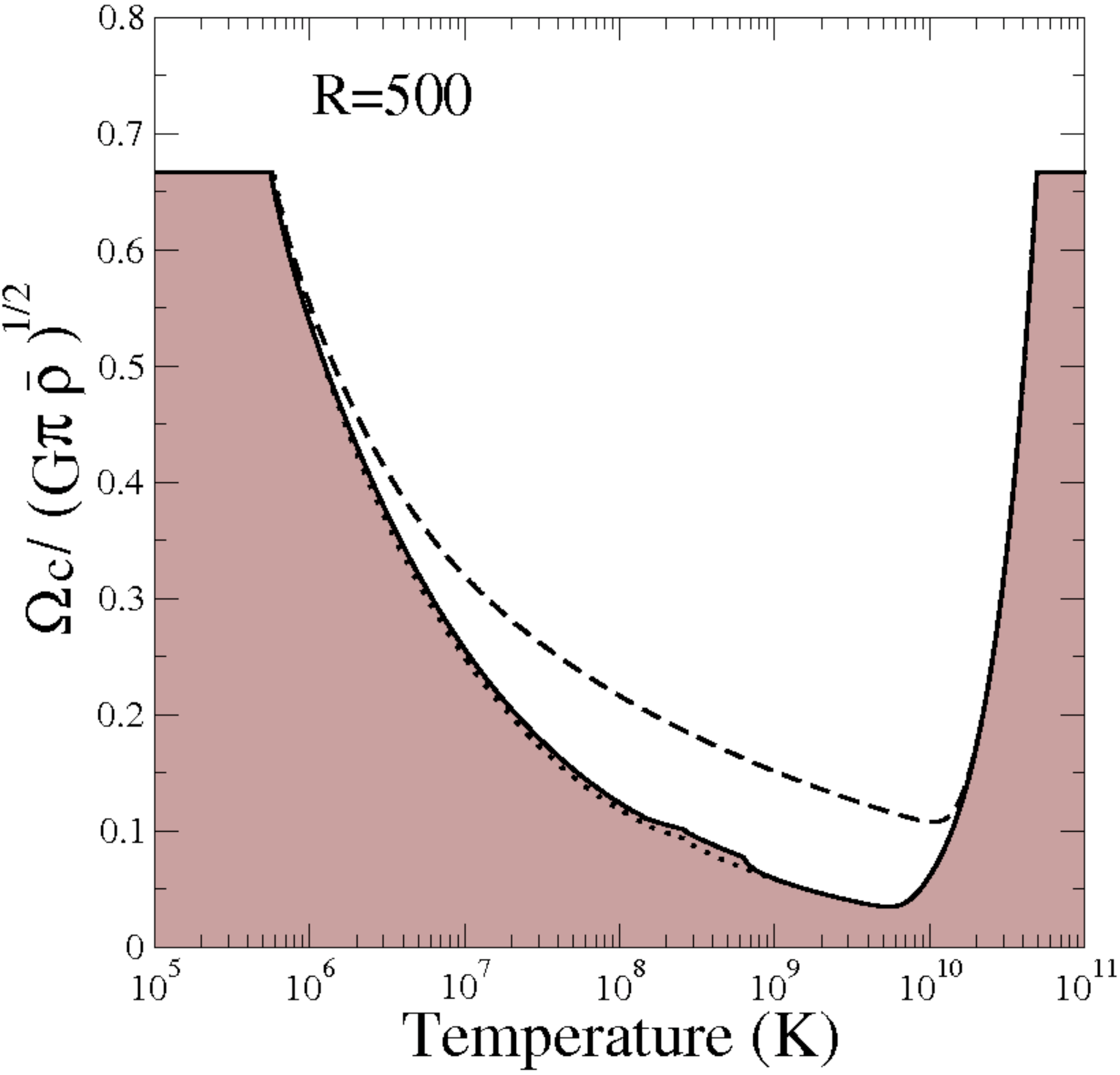}}
\caption{Same as figure~\ref{sw}, but for the ``weak'' superfluidity model.}
\label{sw2}
\end{figure}

The obtained results for the instability window are shown in figures \ref{swr1}, \ref{sw} and \ref{sw2} for a range of drag parameters in the $\mathcal{R}\approx 1$ regime, and the ``strong" and  ``weak" superfluidity models.
The data corresponds to a $M=1.4M_{\odot}$ and $R=10$ km neutron star.
In the figures, we express the critical angular velocity in terms of $\Omega_0=\sqrt{\pi G\bar{\rho}}$ where $\bar{\rho}$ is the average density, and assume that the star cannot spin faster than the breakup limit, taken to be $\Omega_b\approx 2/3\Omega_0$.

Note that, as we are considering a purely fluid star and do not account explicitly for the elastic crust we have not included the damping timescale due to the Ekman layer which is expected to form at the crust-core boundary.
To indicate the effect that the Ekman layer may have on the instability window, we use the rough estimate
\be
\tau_\mathrm{Ek}=3\times 10^5\left(\frac{T}{10^9 \mathrm{K}}\right)\left(\frac{P}{1 \mathrm{ms}}\right)^{1/2}\ \mathrm{s}
\ee
We arrive at this estimate by taking the simple constant density estimate of \citet{review} for a $M=1.4M_{\odot}$ and $R=10$ km neutron star, corrected for a ``slippage'' factor $\mathcal{S}_c$=0.05, as defined by \citet{eck2}. In fact, it has been shown by \citet{eck1} that one should expect the constant density estimate to only differ by factors of a few from the result for a stratified model. Hence, it should be a reasonable approximation for our discussion.

Perhaps not very surprisingly, our analysis demonstrates that the mutual friction
damping can have a significant effect on the r-mode instability window in the extreme case $\mathcal{R}\approx 1$. It is, however, interesting to
note the impact on the instability window of the two superfluidity models.
The results in figures \ref{sw} and \ref{sw2} show that, for values of the drag parameter in the range $0.005 < \mathcal{R} < 500$, mutual friction may have a
 significant effect on the r-mode instability. 
Comparing the two figures, we see that  the "strong" or "weak" superfluidity models lead to significantly different results in the strong friction regime
(keeping the mass and radius of the star fixed). As the star cools below the superfluid transition
temperature the mutual friction timescale initially decreases (the critical frequency increases).
However, at lower temperatures the damping time may increases again (around $T\approx 10^9 K$). This would lead to
a second minimum in the instability curve. For parameter values outside the range $0.005 < \mathcal{R} < 500$, the mutual 
friction would not be the dominant dissipation mechanism for the superfluid r-modes.

It should be stressed that we have focussed on cases where
both $\mathcal{B}^{'}$ and $\mathcal{B}$ are relevant. In the very strong drag regime (i.e. $ \mathcal{R} \gg 1$) one will have $\mathcal{B}^{'}\approx 1$ but $\mathcal{B}\ll 1$. 
In this limit, one again finds that the  mutual friction damping of the r-modes is irrelevant.

\section{Concluding remarks}

In this paper we have re-examined the problem of mutual friction
damping in rotating superfluid neutron stars and its effect on the r-mode instability.
Our analysis differed from previous efforts in that we did not a priori assume that the drag on the
superfluid vortices is weak. By including the mutual friction force in the equations of motion, we were able to
 (for the first time) consider r-modes in the strong drag regime. We  calculated these modes
to second order in rotation, thus  extending the first order results of \citet{fmode}.

Our mode analysis focussed on solutions that are purely axial to leading order. For the
classical  r-mode, which is expected to lead to the fastest growing gravitational-wave instability,
this assumption has the advantage that the equations for the co-moving
degree of freedom decouple and serve as source for
the counter-moving motion in the superfluid region of the star. For the particular case of
an $n=1$ polytrope, and in the Cowling approximation,
we determined a useful analytic solution for the co-moving motion.
The existence of this solution, which is relevant also for the single fluid barotropic r-mode,
simplified the solution of the superfluid problem considerably.

In the superfluid problem there may also exist a class of counter-moving r-modes.
These modes are, however, in general not purely axial to leading order. For a stratified
neutron star model, the velocity field of these modes acquire a leading order polar component.
Such modes cannot be determined within our current framework. This means that we cannot confirm
the proposed existence of "avoided crossings" between the two classes of modes. Such crossings have been
suggested as the explanation for the sharp resonances
with very short mutual friction damping time scales found by \cite{Lind3} and \cite{yl2}.
Our results do not show such resonances. Whether this is an artefact of our approximation scheme is not clear.
We see no evidence that our approximation fails for  particular values of the entrainment,
as one might expect if the r-modes change character as the entrainment is varied.
This issue requires further attention.

We have, however, calculated the counter-moving r-mode in the only case where it can exist, 
when the star is not stratified, see
Appendix~B. In this case it
is possible to obtain, once again, an analytic solution. This solution, which was presented
here for the first time, has already been used as a test bench for numerical
simulations of superfluid neutron stars \citep{Andrea}.

For the classical (mainly co-moving) r-mode, our results confirm that mutual friction due
to the most commonly considered mechanism (electron scattering
off the vortex array) acts on a timescale that is too long to suppress the r-mode instability.
We conclude that, in the weak drag regime, the mutual friction is not the leading damping mechanism
for these modes. This agrees with the conclusions of \cite{Lind3} and \cite{yl2}.

We have also considered, for the first time, the effect of the mutual friction on r-modes in the
strong drag regime. In the extreme limit, this leads to the expected result that
the damping timescale is (again) very long. Hence, mutual friction does not damp r-modes
effectively in the $\mathcal{R}\gg 1$ regime. In order to understand the effect that a strong mutual friction
may have, we have also studied the intermediate regime where $\mathcal{R}\approx 1$,
when $\mathcal{B}^{'}\approx \mathcal{B} \approx 1/2$. This regime could be relevant if there exists
regions in the star where the  interaction between neutron vortices and fluxtubes is efficient \citep{strong1,strong2,strong3}.
In this case we find that mutual friction can have a distinct effect on the r-mode instability window.
The instability may, in fact, be completely suppressed below the superfluid transition temperature.
Moreover, the mutual friction damping depends strongly on the model for the superfluid energy gap,
which determines the critical temperature below which neutrons or protons become superfluid.
We have compared results for two typical models, taken from \citet{nuclphys},
representing "strong" and "weak" superfluidity, respectively. This analysis lays the
foundation for studies of realistic neutron star models, where the size of the superfluid regions
depends explicitly on the stars temperature.

With this analysis we have prepared the ground for a more detailed study of neutron stars with exotic cores,
e.g. dominated by hyperons or deconfined quarks. In each of these cases, one would expect superfluidity
to be relevant. In the case of hyperons, the problem is likely to require additional ``fluid'' degrees of freedom.
Naively, one would expect the neutron and proton superfluid/superconducting mixture to coexist with a neutral $\Lambda$
superfluid and a $\Sigma^-$ superconductor. Assuming electromagnetic coupling between the charged components, this
would leave three distinct hydrodynamical degrees of freedom (at zero temperature). Neutron stars with 
deconfined quark cores may also require additional degrees of freedom. Perhaps the simplest
possibility corresponds to the so-called CFL phase \citep{Alford1999} for which, at low temperatures, one has to consider 
a superfluid coupled to phonon excitations \citep{Mannarelli}. This should lead to a dynamical problem that is formally equal 
to superfluid He$^4$, see \cite{helium} for a recent discussion. More complicated phases, requiring
the inclusion of Kaons, either as thermal excitations or as a condensate, will also need to be 
considered. It would be very interesting, and perhaps important, to understand better to 
what extent multifluid dynamics plays a role for such exotic systems. 

\section*{Acknowledgements}

This work was supported by STFC in the UK through grant number PP/E001025/1.

\appendix

\section{Deriving the r-mode equations}
\label{AppA}

In this Appendix we provide the derivation of the equations that determine the rotational corrections to the r-modes.
We take as our starting point the general slow-rotation perturbation equations from Section~\ref{slowpert}.
Focussing on modes that are purely toroidal to leading order, we assume that the
  total displacement vector takes the form
\be
\frac{\xi_+^i}{a}=\left(0,\frac{K_{lm}}{\sin\theta}\frac{\partial}{\partial\phi}, -K_{lm}\frac{\partial}{\partial\theta}\right)Y^m_l+\sum_{\nu,\mu}\left(S_{\nu\mu},Z_{\nu\mu}\frac{\partial}{\partial\theta},\frac{Z_{\nu\mu}}{\sin{\theta}}\frac{\partial}{\partial\phi}\right)Y^\mu_\nu
\ee
Meanwhile, the difference displacement vector can be written
\be
\frac{\xi_-^i}{a}=\left(0,\frac{k_{lm}}{\sin\theta}\frac{\partial}{\partial\phi}, -k_{lm}\frac{\partial}{\partial\theta}\right)Y^m_l+\sum_{\nu,\mu}\left(s_{\nu\mu},z_{\nu\mu}\frac{\partial}{\partial\theta},\frac{z_{\nu\mu}}{\sin{\theta}}\frac{\partial}{\partial\phi}\right)Y^\mu_\nu
\ee
We use uppercase variables for the co-moving degree of freedom and the corresponding lowercase variable for the counter-moving
degree of freedom.

Our main interest is in the ``classical'' $l=m$ r-mode, a mode that to first order in rotation involves only the co-moving degree of freedom and which is purely toroidal.
This leads to the requirement that the leading term $K_{lm}$ is of order unity while the amplitude of the spheroidal components is of order $\Omega^2$ for the total displacement.
Moreover, the first order results of \citet{fmode} show that all components of the difference
displacement must be of order $\Omega^2$.
With this ordering for the displacement vector, and for the frequency given in equation (\ref{freqord}),  we have the equations
\bear
\sum_{\nu,\mu}a\frac{dS_{\nu\mu}}{da}Y^\mu_\nu&=&\sum_{\nu,\mu}\left[\left(\frac{V}{\Gamma}-3\right)S_{\nu\mu}-\frac{V}{\Gamma_1}\zeta_{\nu\mu}+\nu(\nu+1)Z_{\nu\mu}\right]Y^\mu_\nu\nonumber\\
&&+3imK_{lm}\left(3D_2+a\frac{dD_2}{da}\right)\cos\theta Y^m_l\label{inizio}\\
\sum_{\nu,\mu}a\frac{d\zeta_{\nu\mu}}{da}Y^\mu_\nu&=&\sum_{\nu,\mu}\left[(1-U)\zeta_{\nu\mu}-\frac{V}{\Gamma}x_\p\tau_{\nu\mu}\right]Y^\mu_\nu-2ic_1\omega\tilde{\omega} K_{lm}\sin\theta\frac{\partial Y^m_l}
{\partial\theta}
\eear
These relations follow from the equations for the total displacement,  (\ref{Eulav}) and (\ref{conav}).
From the corresponding equations for the difference, i.e., (\ref{Euldi}) and (\ref{condi}), we get
\bear
\sum_{\nu,\mu}a\frac{ds_{\nu\mu}}{da}Y^\mu_\nu&=&\sum_{\nu,\mu}\left[\left(X-3\right)s_{\nu\mu}-\zeta_{\nu\mu}\frac{1}{(1-x_\p)}\frac{V}{\Gamma}+\nu(\nu+1)z_{\nu\mu}-CS_{\nu\mu}\right]Y^\mu_\nu\label{contro1}\\
\sum_{\nu,\mu}a\frac{d\tau_{\nu\mu}}{da}Y^\mu_\nu&=&
\sum_{\nu,\mu}\left[ (1-U)\tau_{\nu\mu}Y^\mu_\nu\!-2
k_{lm}c_1\omega\tilde{\omega}\left(i\bar{\mathcal{B}}^{'}\sin\theta\frac{\partial Y^m_l}{\partial\theta}\!-\!m\bar{\mathcal{B}}\cos\theta Y^m_{l}\right)-2c_1\omega\tilde{\omega}z_{\nu\mu}\left(m\bar{\mathcal{B}}^{'}+i\bar{\mathcal{B}}\sin\theta\cos\theta\frac{\partial Y_\nu^\mu}{\partial\theta}\right) \right.\nonumber\\
&&\left. +c_1\omega s_{\nu\mu}\left(\omega(1-\bar{\varepsilon})Y^\mu_\nu+2i\tilde{\omega}\bar{\mathcal{B}}(\cos^2\theta-1)Y^\mu_\nu\right) \right]\label{fine}
\eear
In writing down these equations, we have defined the variables 
\bear
\sum_{\nu,\mu}\zeta_{\nu\mu}Y^\mu_\nu&=&\frac{1}{ga}\left(\frac{\delta p}{\rho}\right) \\
\sum_{\nu,\mu}\tau_{\nu\mu} Y^\mu_\nu&=&\frac{1}{ga}\delta\beta
\eear
The various background quantities are defined by equations (\ref{Mdef})--(\ref{epsdef}).

Combining the $\theta$ and $\phi$ components of the co-moving Euler equation we also obtain two algebraic relations
\be
\sum_{\nu,\mu}\nu(\nu+1)\zeta_{\nu\mu}Y^\mu_\nu+2ic_1\omega\tilde{\omega}K_{lm}\left[l(l+1)\cos\theta Y^m_l+\sin\theta\frac{\partial Y^m_l}{\partial\theta}\right]=0\label{inizio2}
\ee
and
\bear
&&K_{lm}\omega\left\{l(l+1)(1+2\epsilon)(\omega-\omega_0)Y^m_l+6D_2\left[\omega\cos\theta\sin\theta\frac{\partial Y^m_l}{\partial\theta}-m\tilde{\omega}(5\cos\theta^5-1)Y^m_l\right]\right\}\nonumber\\
&&-2i\omega\tilde{\omega}\sum_{\nu,\mu}\left\{\left[2S_{\nu\mu}-\nu(\nu+1)Z_{\nu\mu}\right]\cos\theta Y^\mu_\nu+(S_{\nu\mu}-Z_{\nu\mu})\sin\theta\frac{\partial Y^\mu_\nu}{\partial\theta}\right\}=0
\eear
where
\be
\omega_0=\frac{2m\tilde{\omega}}{l(l+1)}
\ee
Analogously, we obtain from the difference Euler equations
\bear
&&\sum_{\nu,\mu}\nu(\nu+1)\tau_{\nu\mu}Y^\mu_\nu=-2ic_1\bar{\mathcal{B}}^{'}\omega\tilde{\omega}k_{lm}\left[l(l+1)\cos\theta Y^m_l+\sin\theta\frac{\partial Y^m_l}{\partial\theta}\right]+2m\bar{\mathcal{B}}c_1\omega\tilde{\omega}k_{lm}\left[2\cos\theta Y^m_l+\sin\theta\frac{\partial Y^m_l}{\partial\theta}\right]\nonumber\\
&&+ c_1\omega \sum_{\nu,\mu} z_{\nu\mu}\left[\omega\nu(\nu+1)(1-\bar{\varepsilon})Y^\mu_\nu-2\tilde{\omega} m\bar{\mathcal{B}}^{'} Y^\mu_\nu-2i\tilde{\omega}\bar{\mathcal{B}}(m^2Y^\mu_\nu+\nu(\nu+1)\cos^2\theta Y^\mu_\nu+2\sin\theta\cos\theta\frac{\partial Y^\mu_\nu}{\partial\theta})\right]\nonumber\\
&&-2c_1\omega\tilde{\omega} \sum_{\nu,\mu}
s_{\nu\mu}\left\{m\bar{\mathcal{B}}^{'}Y^\mu_\nu-i\bar{\mathcal{B}}\left[\cos\theta\sin\theta\frac{\partial Y_\nu^\mu}{\partial\theta}+(3\cos^2\theta-1) Y_{\mu_\nu}\right]\right\}
\eear
and 
\bear
&&k_{lm}\omega l(l+1)(1+2\varepsilon)(1-\bar{\varepsilon})(\omega-\omega_1)Y^m_l
=2i\omega\tilde{\omega}\sum_{\nu,\mu}
\left[(2\bar{\mathcal{B}}^{'}-im\bar{\mathcal{B}})s_{\nu\mu}-
\bar{\mathcal{B}}^{'}\nu(\nu+1)z_{\nu\mu}\right]\cos\theta Y^\mu_\nu\nonumber\\
&&+ \sum_{\nu,\mu} \left[ \bar{\mathcal{B}}^{'}s_{\nu\mu}-(\bar{\mathcal{B}}^{'}-
im\bar{\mathcal{B}})z_{\nu\mu}\right] \sin\theta\frac{\partial Y^\mu_\nu}{\partial\theta}
\label{fine2}
\eear
where
\be
\omega_1=\frac{\tilde{\omega}}{l(l+1)(1-\bar{\varepsilon})}\left\{2m\bar{\mathcal{B}}^{'}+2i\bar{\mathcal{B}}
\left[l(l+1)-m^2\right]\right\}
\ee

We can now use the standard recurrence relations
\bear
\sin\theta\frac{\partial Y^m_l}{\partial\theta}&=&l Q_{l+1} Y^m_{l+1}-(l+1)Q_l Y^m_{l-1}\\
\cos\theta Y^m_l&=&Q_{l+1}Y^m_{l+1}+Q_l Y^m_{l-1}
\eear
with $Q_l$ defined in (\ref{Qdef}).
Repeated use of these relations leads to
\bear
\cos\theta\sin\theta\frac{\partial Y^m_l}{\partial\theta}&=&[lQ^2_{l+1}-(l+1)Q^2_l]Y^m_l+lQ_{l+1}Q_{l+2}Y^m_{l+2}\nonumber\\
&&-(l+1)Q_lQ_{l-1}Y^m_{l-2}\\
\cos^2\theta
Y^m_l&=&(Q^2_{l+1}+Q_l^2)Y^m_l+Q_{l+1}Q_{l+2}Y^m_{l+2}+Q_lQ_{l-1}Y^m_{l-2}
\eear

This way we arrive at the equations that that determine the next order slow-rotation correction to the $l=m$ r-mode;
\bear
a\frac{dS_{l+1}}{da}&=&\left(\frac{V}{\Gamma}-3\right)S_{l+1}-\frac{V}{\Gamma}\zeta_{l+1}+(l+1)(l+2)Z_{l+1}+3imQ_{l+1}
K_{lm}\left(3D_2+a\frac{dD_2}{da}\right)\\
a\frac{d\zeta_{l+1}}{da}&=&(1-U)\zeta_{l+1}-\frac{V}{\Gamma}x_\p\tau_{l+1}-2ic_1\omega\tilde{\omega} l Q_{l+1}K_{lm}\\
a\frac{ds_{l+1}}{da}&=&\left(X-3\right)s_{l+1}-\zeta_{l+1}\frac{1}{x_\p(1-x_\p)}\frac{V}{\Gamma}+(l+1)(l+2)z_{l+1}-CS_{l+1}+3imQ_{l+1}k_{lm}\left(3D_2+a\frac{dD_2}{da}\right)\\
a\frac{d \tau_{l+1}}{da}&=&(1-U)\tau_{l+1}-2c1\omega\tilde{\omega}Q_{l+1}k_{lm}\left(il\bar{\mathcal{B}}^{'}-m\bar{\mathcal{B}}\right)-2c_1\omega\tilde{\omega}z_{l+1}\left\{m\bar{\mathcal{B}}^{'}+i\bar{\mathcal{B}}\left[(l+1)Q_{l+2}^2-(l+2)Q_{l+1}^2\right]\right\}\nonumber\\
&&+c_1\omega s_{l+1}\left\{\omega(1-\bar{\varepsilon})+2i\tilde{\omega}\bar{\mathcal{B}}\left[(Q_{l+2}^2+Q_{l+1}^2)-1\right]\right\} \\
\zeta_{l+1}&=&-2i\omega\tilde{\omega}c_1\frac{l}{l+1}Q_{l+1}K_{lm}\\
\tau_{l+1}(l+1)(l+2)&=&-2i\bar{\mathcal{B}}^{'}\omega\tilde{\omega}c_1{l}{(l+2)}Q_{l+1}k_{lm}+2{m}(l+2)\bar{\mathcal{B}}\omega\tilde{\omega}c_1Q_{l+1}k_{lm}\nonumber\\
&&+c_1\omega z_{l+1}\left\{\omega(l+1)(l+2)(1-\bar{\varepsilon})-2m\tilde{\omega}\bar{\mathcal{B}}^{'}-2i\tilde{\omega}\bar{\mathcal{B}}\left[m^2+Q_{l+2}^2(l+1)(l+4)+Q_{l+1}^2(l+2)(l-1)\right]\right\}\nonumber\\
&&-2c_1\omega\tilde{\omega}s_{l+1}\left\{m\bar{\mathcal{B}}^{'}-i\bar{\mathcal{B}}\left[Q_{l+2}^2(l+4)-Q_{l+1}^2(l-1)-1\right]\right\}\\
\eear
where
\bear
K_{lm}&=&-\frac{i\omega_0}{me}lQ_{l+1}\left[S_{l+1}+(l+2)Z_{l+1}\right]\\
k_{lm}&=&\frac{\omega_0}{m\eta}\left(m\bar{\mathcal{B}}-il\bar{\mathcal{B}}^{'}\right)Q_{l+1}\left[s_{l+1}+(l+2)z_{l+1}\right]
\eear
In the last two expressions we have used
\bear
e&=&(\omega-\omega_0)+3D_2\left\{\frac{2\omega}{l+1}Q_{l+1}^2-\omega_0\left[5Q_{l+1}^2-1\right]\right\}\label{nome}\\
\eta&=&(1-\bar{\varepsilon})(\omega_0-\omega_1)
\eear
 The above equations are identical (\ref{esse})--(\ref{etadef}) in Section~\ref{rpert}.
They completely specify the r-mode problem at second order slow-rotation.

\section{The counter-moving r-mode}

In this appendix, we will examine the ``superfluid'' r-mode, a mode that to first order in rotation involves only the counter-moving degrees of freedom and which is purely toroidal.
This leads to the requirement that the leading term $k_{lm}$ is of order unity while the amplitudes of the spheroidal components are of order $\Omega^2$  for the difference displacement. 
Moreover, all the components of the total displacement are of order $\Omega^2$.
We have already mentioned that such a mode cannot exist unless the proton fraction is constant \citep{fmode}.
Nevertheless, these modes may be of some interest. Hence, it is worth 
determining the relevant slow-rotation corrections.

Given the ordering for the displacement vector, together with the frequency from (\ref{freqord}), we have
the following equations (for the $l=m$ mode)
\bear
a\frac{dS_{l+1}}{da}&=&\left(\frac{V}{\Gamma}-3\right)S_{l+1}-\frac{V}{\Gamma}\zeta_{l+1}+(l+1)(l+2)Z_{l+1}\label{essecontro}\\
a\frac{d\zeta_{l+1}}{da}&=&(1-U)\zeta_{l+1}-\frac{V}{\Gamma}x_\p\tau_{l+1}-2ic_1\omega\tilde{\omega} l Q_{l+1} K_{lm\label{dzc}}\\
a\frac{ds_{l+1}}{da}&=&\left(X-3\right)s_{l+1}-\eta_{l+1}+(l+1)(l+2)z_{l+1}-C S_{l+1}+3imQ_{l+1} k_{lm}\left(3D_2+a\frac{dD_2}{da}\right)\\
a\frac{d \tau_{l+1}}{da}&=&(1-U)\tau_{l+1}-2c_1\omega\tilde{\omega}Q_{l+1}k_{lm}\left(il\bar{\mathcal{B}}^{'}-m\bar{\mathcal{B}}\right)\label{dtaucontro}\\
\zeta_{l+1}&=&-2i\omega\tilde{\omega}c_1\frac{l}{l+1}Q_{l+1}K_{lm}+c1z_{l+1}\left[\omega-\frac{2}{(l+1)(l+2)}\tilde{\omega}\right]-\frac{2m}{(l+1)(l+2)}c_1\omega\tilde{\omega}S_{l+1}\label{orderzetacontro}\\
\tau_{l+1}&=&-2i\bar{\mathcal{B}}^{'}\omega\tilde{\omega}c_1\frac{l}{(l+1)}Q_{l+1}k_{lm}+2\frac{m}{(l+1)}\bar{\mathcal{B}}\omega\tilde{\omega}c_1Q_{l+1}k_{lm}\nonumber\label{taucontro}
\eear
and
\bear
K_{lm}&=&-\frac{i\omega_0}{me}lQ_{l+1}\left[S_{l+1}+(l+2)Z_{l+1}\right]\\
k_{lm}&=&\frac{\omega_0}{m\eta}\left(m\bar{\mathcal{B}}-il\bar{\mathcal{B}}^{'}\right)Q_{l+1}\left[s_{l+1}+(l+2)z_{l+1}\right]
\eear
where
\bear
e&=&(\omega-\omega_0)\label{defe}\\
\eta&=&(1-\bar{\varepsilon})(\omega_0-\omega_1)+3D_2\left\{\frac{2\omega(1-\bar{\varepsilon})}{l(l+1)}\left[lQ_{l+1}^2-(l+1)Q_l^2\right]-\omega_0\left[5Q_{l+1}^2+5Q_l^2-1\right]\right\}
\eear
and we have defined the variable
\bear
\sum_{\nu,\mu}\eta_{\nu\mu} Y^\mu_\nu&=&\frac{1}{x_\p(1-x_\p)}\delta x_\p
\eear

We can now show that one cannot in general have a pure countermoving r-mode, as in this case one has $k_{lm}$ of order unity and thus, from equation (\ref{taucontro}) one finds that $\tau_{lm}$ is of order $\Omega^2$. This means that in equation (\ref{dzc}) the terms involving $\zeta_{lm}$ must be of the same order, but this would lead, from equation (\ref{orderzetacontro}) to one or more components of the total displacement vector being of order unity. The mode would thus no longer be a pure counter-moving r-mode, but a more general inertial mode.
However, this argument fails if we assume constant density profile. In this case there is no coupling term in equation (\ref{dzc}) and one can solve the problem by defining an enthalpy-type variable, see \citet{fmode}.

Hence, we consider an equation of state for which the two fluids are decoupled. In particular, we shall consider the analytical equation of state  of \citet{prix,Andrea}. Then the energy functional takes the form
\begin{equation}
\mathcal{E} = \frac{1}{2} A_{\X\Y} \rho_\X \rho_\Y \, ,
\label{eq:EoS1}
\end{equation}
where $A_{\X\Y}$ is given by
\begin{equation}
A_{\n\n} = \frac{2 K}{1-\left( 1 + \sigma \right) x_\p} \, , \qquad \qquad
A_{\p\p} = \frac{2 K  \left[ 1 + \sigma - \left( 1 + 2 \sigma \right) x_\p \right] } {x_\p \left[  1-\left( 1 + \sigma \right) x_\p \right] }
\, , \qquad \qquad A_{\n \p} = - \sigma A_{\n \n} \, , \label{eq:AxyPCA}
\end{equation}
and $x_\p$, $\sigma$ and $K$ are
constants.
In the background one has $p=K\rho^2$ and a constant proton fraction, while for the perturbations one finds that
\begin{equation}
\delta P     = 2 K \rho \, \delta \rho \, , \qquad \qquad
\delta \beta = \frac{ 2 K \left( 1+\sigma \right) }
                { x_\p \left[ 1 -\left( 1 + \sigma \right) x_\p \right] }
\rho \delta x_\p \, .
\end{equation}
We now see that the coupling term in equation (\ref{condi}) vanishes, as $x_\p$ is constant, and there is also no coupling from the equation of state as i) the pressure is a function of the density only, and ii) $\beta$ is a function of $x_\p$ only. 

Let us consider the inviscid problem. This is of immediate interest since we can test the solution against the 
recent time-evolutions carried out by \citet{Andrea}.
 Neglecting the mutual friction, the equations for the mode we are interested in take the form
\beq
a\frac{ds_{l+1}}{da}&=&(X-3)s_{l+1}+(l+1)(l+2)z_{l+1}+\frac{V}{\Gamma}E\tau_{l+1}\label{con1}\\
a\frac{d\tau_l+1}{da}&=&(1-U)\tau_{l+1}-2ic_1\omega\tilde{\omega}Q_{l+1}lT_{l}\label{con2}
\eeq
where
\be
E=\frac{1+x_\p(1+\sigma)}{(1-x_\p)(1+\sigma)}
\ee
The equations for the co-moving degree of freedom are completely decoupled and take the same form as in equations (\ref{zetan}) and (\ref{eqesse}).
The frequency of the counter-moving r-mode can be written as
\be
\sigma^s=\sigma^s_0\Omega+\sigma^s_2\Omega^3
\ee
with \citep{fmode}
\be
\sigma^s_0=\frac{1}{(1-\bar{\varepsilon})}\frac{2m}{l(l+1)}
\ee
In order to determine the correction $\sigma^s_2$ we adopt the same strategy as for the classical r-mode and use the solution to equation (\ref{con1}) to find a particular solution to equation (\ref{con2}). Skipping the details of the calculation, which is essentially the same as in section \ref{analytics}, we obtain the relation
\be
\sigma^s_2=-\frac{16}{\pi^2}\frac{l}{(l+1)^4(2l+3)}\left[\frac{\pi^{2l+5}}{2(2l+5)}E
-\frac{5\pi^2(l+1)^2}{8}f_2(\pi)\right]\frac{1}{\mathcal{I}_2(1-\bar{\varepsilon})^2}
\ee
Here we have assumend that the entire  star is superfluid, and imposed the boundary condition $\Delta\beta=0$ at the surface, in order to be able to compare 
directly to the numerical results of \citet{Andrea}. This comparison leads to a very good agreement (see figure 2 of \citet{Andrea}).

\end{document}